# AN EXPERIMENT WITH DENOTATIONAL SEMANTICS

(a working version)

DOI: 10.13140/RG.2.2.31272.42249

Andrzej Jacek Blikle

Warsaw, April 16th, 2019

# Abstract

The paper is devoted to showing how to systematically design a programming language in "reverse order", i.e. from denotations to syntax. This construction is developed in an algebraic framework consisting of three many-sorted algebras: of denotations, of an abstract syntax and of a concrete syntax. These algebras are constructed in such a way that there is a unique homomorphism from concrete syntax to denotations, which constitutes the denotational semantics of the language.

Besides its algebraic framework, the model is set-theoretic, i.e. the denotational domains are just sets, rather than Scott's reflexive domains.

The method is illustrated by a layer-by-layer development of a virtual language Lingua: an applicative layer, an imperative layer (with recursive procedures) and an SQL layer where **Lingua** is regarded as an API (Application Programming Interface) for an SQL engine. The latter is given a denotational semantics as well.

Mathematically the model is based on so-called naive denotational semantics [17], many-sorted algebras [21], equational grammars [6], and a three-valued predicate calculus based on a three-valued proposition calculus of J. McCarthy [25]. Three-valued predicates provide an adequate framework for error-handling mechanisms and also for the development of a Hoare-like logic with clean-termination [10] for **Lingua**. That logic is used in [16] for the development of correctness-preserving programs' constructors. This issue is, however, not covered by the paper.

The langue is equipped with a strong typing mechanism which covers basic types (numbers, Booleans, etc.), lists, arrays, record and their arbitrary combinations plus SQL-like types: rows, tables and databases. The model of types includes SQL integrity constraints.

# Contents

# 1 Introduction

## 1.1   Reversing the traditional order of things

The problem of mathematically-provable program-correctness appeared for the first time in a work of Alan Turing [29] published in conference-proceedings *On High-Speed Calculating Machines*, which took place at Cambridge University in 1949. Later for several decades, that subject was investigated usually as *proving program correctness,* but the developed methods never became everyday tools for software engineers. Finally, these efforts were practically abandoned what has been commented in 2016 by the authors of a monography *Deductive Software Verification* [1]:

*For a long time, the term formal verification was almost synonymous with functional verification. In the last years, it became more and more clear that full functional verification is an elusive goal for almost all application scenarios. Ironically, this happened because of advances in verification technology: with the advent of verifiers, such as KeY, that mostly cover and precisely model industrial languages and that can handle realistic systems, it finally became obvious just how difficult and time-consuming the specification of the functionality of real systems is. Not verification but specification is the real bottleneck in functional verification.*

In my opinion, the failure in constructing a practical system for program validation has had two sources.

The first lies in the fact that in building a programming language we start from syntax and only later — if at all — define its semantics. The second source is somehow similar but concerns programs: we first write a program and only then try to prove it correct.

To build a logic of programs for a programming language, one must first define its semantics on a mathematical ground. Since 1970-ties it was rather clear for mathematicians that such semantics to be "practical" must be *compositional*, i.e., the meaning of a whole must be a composition of the meanings of its parts. Later such semantics were called *denotational* — the meaning of a program is its *denotation* — and for about two decades researchers investigated the possibilities of defining denotational semantics for existing programming languages. Two most complete such semantics were written in 1980 for Ada [4] and for CHILL [18] in using a metalanguage VDM [2]. A little later, but in the same decade, a minor exercise in this field was a semantics of a subset of Pascal written in **MetaSoft** [11], the latter based on VDM.

Unfortunately, none of these attempts resulted in the creation of software-engineering tools that would be widely accepted by the IT industry. In my opinion that was unavoidable since for the existing programming languages a full denotational semantics simply cannot be defined (see Sec.3). That was, in turn, the consequence of the fact that historically syntaxes were coming first and only later researchers were trying to give them a mathematical meaning. In other words — the decision of how to describe things preceded the reflection of what to describe.

In addition to that, two more issues were complicating denotational models of programming languages. They were related to two mechanisms considered important in 1960-ties but later abandoned and forgotten. One was a common *jump instruction* **goto**, the other — specific procedures that may take themselves as parameters (Algol 60, see [26]). The former has led to *continuations* (see [22]), the latter to *reflexive domains* (see [27]). Both contributed to a technical complexity of denotational models which was discouraging not only for practitioners but even for mathematicians.

The second group of problems followed from a tacit assumption that in the development of a mathematically correct program the development of a program should precede the proof of its correctness. Although this order is quite obvious in mathematics — first theorem, then its proof — it is rather awkward for an engineer who first performs all necessary calculations (the proof) and only then builds his bridge or aeroplane.

The idea "first a program and correctness-proof later" seems not only irrational but also practically rather unfeasible for two reasons.

First reason follows from the fact that a proof of a theorem is usually longer than the theorem itself. Consequently, proofs of program correctness should contain thousands if not millions of lines. It makes "hand-made proofs" rather unrealistic. On the other hand, automated proofs were not available by the lack of formal semantics for existing programming languages.

Even more important seem, however, the fact that programs that are supposed to be proved correct are usually incorrect! Consequently, correctness proofs are regarded as a method of detecting errors in programs. In other words, we are first doing things in a wrong way to correct them later. Such an approach does not seem very rational either.

As an attempt to cope with all the mentioned problems I propose some mathematical tools and methods that allow for the development of programming languages with denotational semantics. Their detailed description may be found in a preprinted book. To illustrate these methods an exemplary programming language, **Lingua** has been developed from denotations to syntax (first publication of that method in [12]). In this way, the decision of what to do (denotations) precedes the decision of how to express that (syntax).

Mathematically both the denotations and the syntaxes constitute many-sorted algebras (Sec.2.2), and the associated semantics is the homomorphism from syntax to denotations. As turns out, there is a simple method — to a large extend algorithmizable — of deriving syntax from (the description of) denotations and the semantics from both of them.

At the level of data structures, **Lingua** contains Booleans, numbers, texts, records, arrays and their arbitrary combinations plus SQL databases. It is also equipped with a relatively rich mechanism of types, e.g. covering SQL-like integrity constraints[1], and with tools allowing the user to define his/her own types structurally. At the imperative level, **Lingua** contains structured instructions, type definitions, procedures with recursion and multi-recursion and some preliminaries of object programming.

The issue of concurrency is not tackled in [16] since the development of a "fully" denotational semantics for concurrent programs (if at all possible) would require separate research[2].

Ones we have a language with denotational semantics, we can define program-construction rules that guarantee the correctness of programs developed in using these rules. This method was for the first time sketched in my paper [8] and in [16] is described in Sec.8. It consists in developing so-called *metaprograms* which syntactically include their specifications. The method guarantees that if we compose two or more correct programs into a new program, we get a correct program again. The correctness proof of a program is hence implicit in the way the program has been developed.

---

[1] Except subordination relations which are described by a different mechanism.
[2] There exist mathematical semantics of concurrency which can be said to be only "partially denotational". An example of such a solution is a "component-based semantics" (cf. [2]), where the denotations of programs' components are assigned to programs in a compositional way (i.e. the denotation of a whole is a composition of the denotations of its parts), but the denotations themselves are so called fucons whose semantics is defined operationally.

Basic mathematical tools used in my denotational models are the following:

1. fixed-point theory in partially ordered sets,
2. the calculus of binary relations,
3. formal-language theory and equational grammars,
4. fixed-point domain-equations based on so-called *naive denotational semantics* (cf. [17]),
5. many-sorted algebras,
6. abstract errors as a tool for the description of error-handling mechanisms,
7. three-valued predicate calculi of McCarthy and Kleene,
8. the theory of total correctness of programs with clean termination (cf. [10]).

All these tools are described in Sec.2 and Sec.8 of [16], and some of them are sketched in Sec.1.4 of the present paper.

In constructing **Lingua,** I assume three priorities regarding the choice of programming mechanisms:

- the priority of the simplicity of the model, i.e., the simplicity of denotations, syntax, and semantics; this has laid to the resignation from, e.g., `goto` instruction and self-applicative procedures,
- the priority of the simplicity of program-construction rules; e.g., the assumption that the declarations of variables and procedures, as well as the definitions of types, should be located at the beginning of a program,
- the priority of protection against "oversight errors" of a programmer; e.g., the resignation of global variables in procedures and of side-effects in functional procedures.

All these commitments forced me to give up some programming constructions which — although denotationally definable — would lead to complicated descriptions and even more complicated program-construction rules. It is worth mentioning in this place that the priority of simplicity is not new in the history of programming languages. For that very reason, programming-language designers abandoned `goto`-s as well as self-applicative procedures.

The name **Lingua** has been chosen to commemorate the circumstances under which from October to December 1969 I wrote my first denotational semantics of a very simple programming language (this work was later published in Dissertationes Mathematicae [5] as my habilitation (postdoctoral) thesis). During three months as a scholar of the Italian Government, I was working in the Istituto di Elaborazione dell'Informazione in Pisa. I didn't yet know the works of Dana Scott or the concept of denotational semantics, and I constructed my language and its semantics on a model theory known in mathematical logic. Only eighteen years later, in the year 1987, I described (in [12]) the idea of how to develop syntax from detonations.

## 1.2 What is in the paper

I am deeply convinced that one can talk about programming in a precise and clear way. I also believe that taking responsibility by software engineers should be possible in the same way as it is in the case of the engineers of cars, bridges or aeroplanes. However, I am aware of the fact that the existing tools for software engineers do not allow for the realisation of any of these goals.

The paper contains many thoughts developed in the years 1960-1990 that later have been abandoned. One of the teams developing these ideas was working in the Institute of Computer Science of the Polish Academy of Sciences, and I had the pleasure to chair it. At that time we have developed a semi-formal metalanguage MetaSoft dedicated to formal definitions of programming languages (cf. [11]). This metalanguage is used in [16] and in the present paper as a definitional vehicle for denotational models.

I am aware of the fact that the content of [16] represents a very restricted part of the world of today's programming languages. Something had to be chosen, however, to begin with. **Lingua** contains, therefore, a selection of programming tools that have been known for many years and that are still in use. In the future, I shall try to complete my models with those vehicles that my readers will consider important. I also hope that maybe some of my readers will undertake this challenge. Feel invited to cooperate.

## 1.3    What this paper is not offering

The quality of a program consists in:

1. the compatibility of the program's specification with the expectations of its user,
2. the compatibility of the program itself with its specification.

In this paper, and in [16], I am tackling only the second aspect. My choice is not caused by the fact that the first problem is less important, or that it has been already solved, but only because the second problem was the main subject on my research for two decades and therefore I dare to talk about it now[3].

I also have to emphasise very strongly that my virtual language **Lingua** is not regarded neither as a practical programming language nor even as a standard of such a language although maybe a real language will grow from **Lingua** in the future. At present, it only offers a platform where to explain the constructions and the models discussed in [16]. I have tried to cover in it the selected basic tools that are present in languages which are known to me today. I resigned form concurrency, and object programming is in [16] only roughly sketched.

I believe, however, that there are enough applications today that can be developed in using the tools described in [16].

## 1.4    What is new in my approach

By "my approach" I understand the ideas and techniques described in my early papers from [6] to [15], which have been summarised and extended in the preprint book [16]. All these ideas base on concepts well-known for years:

- denotational semantics of D. Scott's and Ch. Strachey's (cf. [27], [28]),
- generative grammars of N. Chomsky's (cf. [19], [20]),
- Hoare's logic of programs (cf. [23]),

---

[3] I am convinced that the first problem is equally fascinating as the second. I would very much welcomed any initiative of a cooperation in this field.

- on many-sorted algebras introduced to the mathematical foundations of computer science by J. A Goguen, J.W, Thatcher, E.G Wagner and J.B Wright (cf. [21]),
- three-valued propositional calculus S.C. Kleene's (cf. [24]).

What — I believe is new in my approach — is the following:

1. **Programming language design and development**:

    1.1. Denotational model based on set-theory rather than on D. Scott's reflexive domains which makes the model much simpler and easy to be formalized.

    1.2. A model of data-types that covers not only structured and user-defined types but also SQL integrity constraints.

    1.3. A formal, and to a large extend an algorithmic method of a systematic development of syntax from denotations and of a denotational semantics from both of them.

    1.4. The idea of a colloquial syntax which allows making syntax user-friendly without damaging a denotational model.

    1.5. Systematic use of error-elaboration in programs supported by a three-valued predicate calculus.

2. **The development of correct programs**

    2.1. A method of systematic development of correct programs with their specifications, rather than an independent development of programs and specifications followed by program-correctness proof.

    2.2. The use of three-valued predicates to extend Hoare's logic by a clean termination property.

3. **General mathematical tools**

    3.1. Equational grammars applied in defining the syntax of programming languages.

    3.2. A three-valued calculus of predicates applied in designing programming languages and in defining sound program constructors for such languages.

# 2  Mathematical preliminaries

For a full description of mathematical tools used in the development of denotational models see Sec.2 of [16] Below there is a selection of concepts and notations that are used in the present paper. They all come from **MetaSoft** [11] — a metalanguage for the description of programming languages[4].

## 2.1  Notational conventions

I do not assume that the reader is acquainted with [16] and therefore I use only as much of my metalanguage as necessary to make the paper sufficiently clear and concise. Let me start with some basic notations:

- a : A means that a is an element of the set A; according to the denotational dialect sets are most frequently called *domains*,

---

[4] Developed in the decade 1980-1990 in the Institute of Computer Science of the Polish Academy of Sciences by a team which I had a honor to chair.

- f.a denotes f(a), and f.a.b.c denotes ((f(a))(b))(c); intuitively f takes a as an argument and returns the value f(a) which is a function which takes b as an argument and returns the value (f(a))(b), which is again a function…
- f ● g denotes the sequential composition of functions, i.e. (f●g).a = g.(f.a)
- A → B denotes the set of all *partial functions* from A to B, i.e., functions which are (possibly) undefined for some elements of A,
- A ↦ B denotes the set of all *total functions* from A to B, i.e., functions undefined for all elements of A; of course, A ↦ B is a subset of A → B,
- A ⇒ B denotes the set of all finite function from A to B, i.e. functions defined for only finite subsets of A; such functions are called *mappings,* and of course, each mapping is a particular case of a partial function,
- [$a_1$/$b_1$,…,$a_n$,$b_n$] denotes a mapping that assigns $b_i$ to $a_i$ and is undefined otherwise,
- A | B denotes the set-theoretic union of A and B,
- A x B denotes the Cartesian product of A and B,
- $A^{c*}$ denotes the set of all finite (possibly empty) tuples of the elements of A,
- $A^{c+}$ denotes the set of all finite non-empty tuples of the elements of A,
- If L is a formal language (i.e. a set of words), then L* denotes the set of all finite concatenations of words in L,
- tt and ff denote logical values „true" and „false" respectively,
- many-character symbols like dom, bod, com denote metavariables running over domains and if they are written with quotation marks as 'abdsr' denote themselves, i.e., metaconstants[5].
- in the definitional clauses of **Lingua** instead of indexed variables like $sta_1$, we write sta1 or sta-1 which is closer to a notation used in programs.

In this paper three different linguistic levels are distinguished:

1. the level of the basic text of the paper written in Times New Roman,
2. the level of a formal, but not formalized, metalanguage **MetaSoft** written in Arial,
3. the level of formalized programming language **Lingua** whose syntax, i.e. programs are written in `Courier New`.

The difference between "formal" and "formalized" is such that the former is introduced intuitively as a mathematical notation, whereas the latter requires an explicit definition of syntax (usually by a grammar) and a formal definition of semantics.

A frequently used construction in **MetaSoft** is a *conditional definition of a function* with the following scheme:

f.x =
    $p_1$.x ➜ $g_1$.x
    $p_2$.x ➜ $g_2$.x

---

[5] Metavariables and metaconstants are objects of the metalanguage **MetaSoft** whereas variables and constants are objects of the programming language **Lingua**.

...
    **true** ➔ $g_n.x$

where each $p_i$ is a classical predicate, i.e., a total function with logical values tt or ff, **true** is a predicate which is always satisfied, and each $g_i$ is just a function. The formula above is read as follows:

   if $p_1.x$ is true, then $f.x = g_1.x$ and otherwise,

   if $p_2.x$ is true, then $f.x = g_2.x$ and otherwise,

   ...

   and in all other cases $f.x = g_n.x$.

Intuitively speaking the evaluation of such a function goes line by line and stops at the first line where $p_i.x$ is satisfied.

In the scheme above I also allow the situation where, in the place of a $g_i.x$ we have the undefinedness sign "?" which means that for $x$ that satisfies $p_i.x$ the function $f$ is undefined. This convention is used in conditional definitions of partial functions.

In such definitions we also use a technique similar to defining local constants in programs. For instance if $f : A \times B \mapsto C$ we can write

   $f.x =$

     $p_1.x$ ➔ $g_1.x$

     **let**

       $(a, b) = x$

     $p_2.a$ ➔ $g_2.x$

     $p_3.b$ ➔ $g_3.x$.

which is read as: "let $x$ be a pair of the form $(a, b)$". We can also use **let** in the following way:

   $f.x =$

     $p_1.x$ ➔ $g_1.x$

     **let**

       $y = h.x$

     $p_2.x$ ➔ $g_2.y$

     $p_3.x$ ➔ $g_3.y$.

All these explanations are certainly not very formal, but the notation should be clear when it comes to concrete examples in the sequel of the paper.

By $[a_1/v_n,...,a_n/v_n]$ I denote a finite-domain functions with domain $\{a_1,...,a_n\}$ and the corresponding values $\{v_1,...,v_n\}$. By $f[a_1/v_n,...,a_n/v_n]$ I denote an overwriting of $f$ by $[a_1/v_n,...,a_n/v_n]$, i.e. a function which differs from $f$ only on the domain $\{a_1,...,a_n\}$.

For any two functions $f : A \to B$ and $g : B \to C$ by $f \bullet g$ denotes the *sequential composition* of these functions, i.e.

   $(f \bullet g).a = g.(f.a)$

## 2.2 Many-sorted algebras

The denotational model of a programming language investigated in [16] is based on the concept of a *many-sorted algebra*. Half formally, a many-sorted algebra is a finite collection of sets, called the *carriers* or *sorts* of the algebra, and a finite collection of functions called the *constructors* of the algebra. The constructors take arguments from and return their values to carriers. A graphical representation of a two-sorted algebra of numbers and Booleans is shown in Fig. 2.2-1. This algebra will be referred to as NumBool.

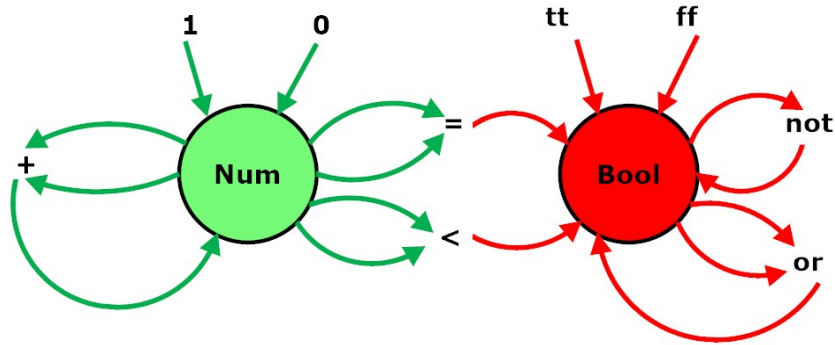

**Fig. 2.2-1 Graphical representation of a two-sorted algebra** NumBool

A textual representation of NumBool — called the *signature* of this algebra — is shown in the left part of Fig. 2.2-2.

| The algebra NumBool | | | The algebra NumBoolExp | | |
|---|---|---|---|---|---|
| 1 | : | $\mapsto$ Num | 1 | : | $\mapsto$ NumExp |
| 0 | : | $\mapsto$ Num | 0 | : | $\mapsto$ NumExp |
| + | : Num x Num | $\mapsto$ Num | + | : NumExp x NumExp | $\mapsto$ NumExp |
| = | : Num x Num | $\mapsto$ Bool | = | : NumExp x NumExp | $\mapsto$ BoolExp |
| < | : Num x Num | $\mapsto$ Bool | < | : NumExp x NumExp | $\mapsto$ BoolExp |
| tt | : | $\mapsto$ Bool | tt | : | $\mapsto$ BoolExp |
| ff | : | $\mapsto$ Bool | ff | : | $\mapsto$ BoolExp |
| not | : Bool | $\mapsto$ Bool | not | : BoolExp | $\mapsto$ BoolExp |
| or | : Bool x Bool | $\mapsto$ Bool | or | : BoolExp x BoolExp | $\mapsto$ BoolExp |

**Fig. 2.2-2 The signatures of two mutually similar algebras**

In our algebra, we have four zero-argument constructors 1, 0, tt, ff, one one-argument constructor not, and four two-argument constructors +, =, <, or. The zero-argument constructors create elements of carriers "from nothing", whereas all other constructors create elements of carriers from other elements of carriers.

An element of an algebra is called *reachable* if it can be constructed (reached) using the constructors of the algebra. In NumBool, where Num denotes the set of <u>all</u> real numbers, the *reachable subset* of Num contains only non-negative integers.

By a *reachable subalgebra* of an algebra we mean its subalgebra with carriers restricted to their reachable parts. In our case, this is an algebra of nonnegative integers and Booleans.

An algebra is said to be *reachable* if all its carriers contain only reachable elements. Notice that if we remove the zero-argument constructor 1 from NumBool, then the reachable subset of Num becomes empty.

In the algebraic approach to denotational models, the algebra of program *denotations* (meanings) is usually unreachable, whereas the algebras of syntax are reachable by definition (see Sec.2.3).

On the right-hand side of Fig. 2.2-2 we have the signature of a syntactic algebra NumBoolExp of (variable-free) expressions. This algebra is *similar* to NumBool in the sense that there is a one-one correspondence between the constructors and the carriers of both algebras, and the "types of constructors" in one algebra are similar to the types in the other (for a formal definition see Sec.2.11 of [16]). In our example this correspondence is implicit row-by-row in the notation: `1` corresponds to 1, `0` corresponds to 0, NumExp corresponds to Num, `+` corresponds to +, etc. The constructors of NumBoolExp create expressions. E.g. the constructor + given two numeric expressions `nexp-1` and `nexp-2` creates the expression[6]:

`+(nexp-1, nexp-2)`

Examples of expressions are:

`1, 0, +(1,1), +(1,+(1,0)), tt, not(<(1,+(1,1))`

We shall assume that NumBoolExp contains only reachable expressions. Such algebra is implicit in the signature of NumBool and, due to its reachability, is unique. Traditionally it is called the *abstract syntax* of the algebra NumBool.

It may be easily proved that for every algebra Alg — and in fact for its signature — there exists a unique algebra of abstract syntax AbsSyn. It is also easy to prove that there exists a unique homomorphism:

   As : AbsSyn $\mapsto$ Alg

We call it the *abstract semantics* of AbsSyn. Of course, a homomorphism between many-sorted algebras is a tuple of functions — one for every carrier. In the case of our example we have two corresponding functions:

   SemE : NumExp $\mapsto$ Num

   SemB : BoolExp $\mapsto$ Bool

which satisfy the equations (called the *semantic clauses*):

   SemE.[`1`] = 1

   SemE.[`+(nexp-1, nexp-2)`] = SemE.[`nexp-1`] + SemE.[`nexp-2`]        (2.2-1)

   SemB.[`<(nexp-1, nexp-2)`] = SemE.[`nexp-1`] < SemE.[`nexp-2`]

   etc.

For instance :

   SemE.[`+(1,+(1,0))`] = 2

   SemB.[`<(+(1,+(1,0)),0)`] = ff

Notice that our homomorphism is "gluing" many different expressions into the same number or Boolean element, e.g.

   SemE.[`+(1,+(1,0))`] = SemE.[`+(1,1)`] = 2

---

[6] For simplicity I use here the same symbol "+" to denote a constructor of expressions and a syntactic symbol of addition.

```
SemB.[<(+(1,+(1,0)),0)] = SemB.[<(0,0)] = ff
```

The notation of an abstract syntax is rather awkward and therefore abstract syntax is usually transformed into a *concrete syntax,* which is more "user-friendly". In our case it would correspond to an infix notation where the concrete + given two expressions `nexp-1` and `nexp-2` creates the expression:

```
(exp-1 + exp-2)
```

and similarly for other constructors. From an algebraic perspective concrete syntax is an algebra — let's denote it by ConSyn — defined in a way that guarantees the existence of two homomorphisms:

　　Co : AbsSyn $\mapsto$ ConSyn　　　　— the concretization of abstract syntax

　　Cs : ConSyn $\mapsto$ Alg　　　　　— the (unique) concrete-syntax semantics.

and moreover that

　　As = Co • Cs

More about a denotational model of programming languages in Sec. 3.2. Readers interested in the mathematical justifications of the model are referred to sections from 2.10 to 2.13 of [16] and to the references given there.

## 2.3　　Equational grammars

Let A be an arbitrary finite set of symbols called an *alphabet*. By a *word* over A, we mean every finite sequence of the elements of A including the empty sequence ε. If p and q are words, then by their *concatenation* — in symbols pq — we mean a sequential combination of these words.

　　Sets of words over A are called *formal languages* or just *languages* over A. If P and Q are languages, then the language

　　PQ = {pq | p : P **and** q : Q}

is called the *concatenation* of P and Q. Similarly to the Cartesian + and * defined in Sec.2.1 we define analogous operations on languages:

　　$P^0$ = {ε}, $P^n$ = $PP^{n-1}$　　for n > 0

　　$P^+$ = U { $P^n$ | n > 0}

　　$P^*$ = $P^+$ | $P^0$

By an *equational grammar* over an alphabet A we mean a set of recursive equations of the form:

　　$X_1$ = $p_1$.($X_1$,…,$X_n$)

　　…

　　$X_n$ = $p_n$.($X_1$,…,$X_n$)

where $X_i$'s run over languages over A and all $p_i$'s are operations on languages constructed as combinations of finite languages (constants), union, concatenation, power, star and plus operations. It may be proved that every equational grammar has a unique least[7] solution which constitutes a tuple ($P_1$,…,$P_n$) of languages. Such a tuple will be called a *many-sorted language*.

---

[7] In the sense of a component wise inclusion.

Every equational grammar defines unambiguously a reachable algebra of words. The following grammar defines the algebra **NumBoolExp** of Sec.2.2:

NumExp = `0` | `1` | `+(`NumExp, NumExp`)`

BoolExp = `tt` | `ff` | `=(`NumExp, NumExp`)` | `<(`NumExp, NumExp`)` |
 `not(`BoolExp`)` | `or(`BoolExp, BoolExp`)`

According to a usual style for writing grammars, the symbols `0`, `1`, `tt`, `ff`, `+`, `=`, `<`, `not`, `or`, `(`, `)` and the coma denote one-element languages: {`0`}, {`1`},…

Equational grammars correspond closely to context-free grammars introduced by Noam Chomsky (e.g. in [19]) in the sense that for each context-free grammar there exists an equational grammar that defines the same many-sorted language, and for a certain class of equational grammars there exists an equivalent context-free grammar. They have been introduced in [6] and are also described in Sec.2.5 and Sec.2.14 of [16].

## 2.4 Abstract errors

For practically all expressions appearing in programs their values in some circumstances cannot be computed "successfully". Here are a few examples:

- the value of x/y cannot be computed if y = 0,
- the value of the expression x+1 cannot be computed if x has not been declared in the program,
- the value of x+y cannot be computed if the sum exceeds the maximal number allowed in the language,
- the value of the array expression a[k] cannot be computed if k is out of the domain of array a, or if a is not an array,
- the query "Has John Smith retired?" cannot be answered if John Smith is not listed in a database.

In all these cases a well-designed implementation should stop the execution of a program and generate an error message or perform a recovery procedure.

To describe that mechanism formally, we introduce the concept of an *abstract error*. In a general case abstract errors may be anything, but in our models, they are going to be texts such as, e.g., 'division-by-zero'. They are closed in apostrophes to distinguish them from metavariables.

The fact that an attempt to evaluate x/0 raises an error message can be now expressed by the equation:

x/0 = 'division-by-zero'

In the general case with every domain **Data**, we associate a corresponding domain with abstract errors

DataE = Data | Error

where **Error** denotes the set of all abstract errors that are generated by our programs. Consequently every partial operation

op : $Data_1$ x … x $Data_n$ → Data

whose partiality is computable[8] may be extended to a total operation

    ope : DataE$_1$ x … x DataE$_n$ ⟼ DataE

Of course **ope** should coincide with **op** wherever **op** is defined.

The operation **ope** is said to be *transparent for errors* or simply *transparent* if the following condition is satisfied:

    if d$_k$ is the first error in the sequence d$_1$,…,d$_n$, then ope.(d$_1$,…,d$_n$) = d$_k$

Intuitively this condition means that arguments of **ope** are evaluated one-by-one from left to right, and the first error (if it appears) becomes the final value of the computation.

The majority of operations on data that will appear in our models are transparent. Exceptions are boolean operations discussed in Sec.2.5

Error-handling mechanisms may be implemented in such a way, that errors serve only to inform the user that (and why) program execution has been aborted. Such a mechanism is called *reactive*. Another option is that the generation of an error results in an action, e.g. of recovering the last state of a database. Such mechanisms are called *proactive*.

A reactive mechanism may be quite easily enriched to a proactive one (see Sec.6.1.8 and Sec.12.7.6.4 of [16]). However, since the latter is technically more complicated, in this paper only reactive model will be discussed.

A well-defined error-handling mechanism allows avoiding situations where programs are aborted without any explanation, or even worse — when they generate an incorrect result without a warning of the user.

## 2.5    Three-valued propositional calculus

Tertium non datur — used to say ancients masters. Computers denied this principle.

In the Aristotelean classical logic, every sentence is either true or false. The third possibility does not exist. However, in the world of computers, the third possibility is not only possible but just inevitable. E.g. in evaluating a boolean expression **x/y>2** an error will appear if **x=0**.

To describe the error-handling mechanism of boolean expressions, we introduce a domain of Boolean values with an error

    BooleanE = {tt, ff, ee}.

In this case, **ee** stands for "error", but in fact, represents either an error or an infinite computation (a looping). In this section, we assume for simplicity that there is only one error. This assumption does not disturb the generality of our model as long as all errors are handled in the same way.

Now, it turns out that the transparency of boolean operators would not be an adequate choice. To see that consider a conditional instruction:

    **if** x ≠ 0 **and** 1/x < 10 **then** x := x+1 **else** x := x−1 **fi**

---

[8] Informally speaking a partiality of a function F is computable if we can write a procedure which given an arbitrary tuple d$_1$,…,d$_n$ of arguments of F will check if F.(d$_1$,…,d$_n$) is or is not defined. E.g. for an array expression arr[k] we can check if the index k belongs to the index range of the array arr. From the general theory of computability we know, however, that there exist functions with non-computable partialities.

We would probably expect that for x=0 one should execute x:=x-1. If however, our conjunction would be transparent, then the expression

x ≠ 0 **and** 1/x < 10

would evaluate to 'division-by-zero' which means that the program aborts. Notice also that the transparency of **and** would imply

ff **and** ee = ee

which would mean that an interpreter that evaluates p **and** q first evaluates both p and q — as in "usual mathematics" — and only later applies **and** to them. Such a mode is called an *eager evaluation*.

An alternative to it is a *lazy evaluation* where, if p = ff, then the evaluation of q is skipped, and the final value of the expression is ff. In such a case:

ff **and** ee = ff

tt **or** ee = tt

A three-valued propositional calculus with lazy evaluation was described in 1961 by John McCarthy (in [25]) who defined boolean operators as shown in Tab. 2.5-1

| **or-m** | tt | ff | ee |
|---|---|---|---|
| tt | tt | tt | tt |
| ff | tt | ff | ee |
| ee | ee | ee | ee |

| **and-m** | tt | ff | ee |
|---|---|---|---|
| tt | tt | ff | ee |
| ff | ff | ff | ff |
| ee | ee | ee | ee |

| **not-m** | |
|---|---|
| tt | ff |
| ff | tt |
| ee | ee |

**Tab. 2.5-1 Propositional operators of John McCarthy**

To see the intuition behind the evaluation of McCarthy's operators consider the expression p **or-m** q assuming that its arguments are computed from left to right[9]:

- If p = tt, then we give up the evaluation of q (lazy evaluation) and assume that the value of the expression is tt. Notice that in this case we maybe avoid an error message or an infinite computation that could be generated by q.

- If p = ff, then we evaluate q, and its value — possible ee — becomes the value of the expression.

- If p = ee, then this means that the evaluation of our expression aborts or loops at the evaluation of its first argument, hence the second argument is not evaluated. Consequently, the final value of the expression must be ee.

The rule for **and** is analogous. Notice that McCarthy's operators coincide with classical operators on classical values (grey fields in the tables). McCarthy's implication is defined classically:

p **implies-m** q = (**not-m** p) **or-m** q

As it turns out, not all classical tautologies remain satisfied in McCarthy's calculus. Among those that are satisfied we have:

---

[9] The suffix "-m" stands for "McCarthy" and is used to distinguish McCarthy's operators not only from classical ones but also from the operators of Kleene, which are used in SQL.

- associativity of **and** and **or**,
- De Morgan's laws

and among the non-satisfied are:

- **or-m** and **and-m** are not commutative, e.g., ff **and-m** ee = ff but ee **and-m** ff = ee,
- **and-m** is distributive over **or-m** only on the right-hand side, i.e.

    p **and-m** (q **or-m** s) = (p **and-m** q) **or-m** (p **and-m** s) however

    (q **or-m** s) **and-m** p ≠ (q **and-m** p) **or-m** (s **and-m** p) since

    (tt **or-m** ee) **and-m** ff = ff  and  (tt **and-m** ff) **or-m** (ee **and-m** ff) = ee

- analogously **or-m** is distributive over **and-m** only on the right-hand side,
- p **or-m** (**not** p) does not need to be true but is never false,
- p **and-m** (**not** p) does not need to be false but is never true.

# 3   General remarks about denotational models

## 3.1    Why do we need denotational models?

Denotational models of programming languages serve as a starting point for the realisation of three tasks:

1. building the implementation of the language, i.e. its parser and interpreter or compiler,
2. creating rules of building correct specified programs,
3. writing a user manual.

In building a language in this way, we should observe one very important (although not quite formal) principle of simplicity:

*A programming language should be as simple and easy to use as possible, although without damaging its functionality, mathematical clarity and the completeness of its description. The same applies to the manual of languages and to the rules of building correct programs.*

This principle shall be realised by caring to make:

1. the syntax of the language as close as possible to the language of intuitive mathematics, for example, whenever this is common, we use infix notation and allow the omission of "unnecessary" parentheses,
2. the structure of the language (i.e. program constructors) leading to possibly simple rules of constructing correct programs (Sec.8 of [16]),
3. the semantics of the language easy to understand by the user rather than convenient for the builder of implementation; for the latter an implementation-oriented equivalent model may be written.

Special attention should be given to point 2 because the simplicity of the rules of building correct programs leads to a better understanding of programs by programmers. This fact was realised already in the years 1970 and has led to the elimination of goto instructions. This decision resulted in a major simplification of programs' structures, which increased their reliability.

Following point 3, I will sometimes — as common in mathematics — "forget" about the difference between syntax and denotations. E.g. I will talk about the value of an expression x + y, rather than about the value of its detonation. I would say that the instruction `x:=y+1` modifies variable x, instead of saying that the denotation of this instruction modifies the memory state at variable x, etc. Of course, on a formal level syntax will be precisely distinguished from denotations.

## 3.2   Five steps to a denotational model

Building up **Lingua** I refer to an algebraic model described in Sec.2.2. This model corresponds to the diagram of three algebras shown in Fig. 3.2-1. We build it in such a way that the existence of the semantics Cs of concrete syntax is insured, and the equation:

   As = Co ● Cs

is satisfied.

The construction of a denotational model begins with an algebra of detonation Den. Its constructors unambiguously determine the reachable subalgebra ReDen. From the signature of Den, we unambiguously derive the abstract syntax algebra AbsSy. The first of these steps is creative since it comprises all the major decisions about the future language. Contrary to it, the derivation of AbsSy can be performed algorithmically. The corresponding algorithm takes the description — e.g. in **MetaSoft —** of the signature of Den. This technique will be explained in more details in the subsequent sections.

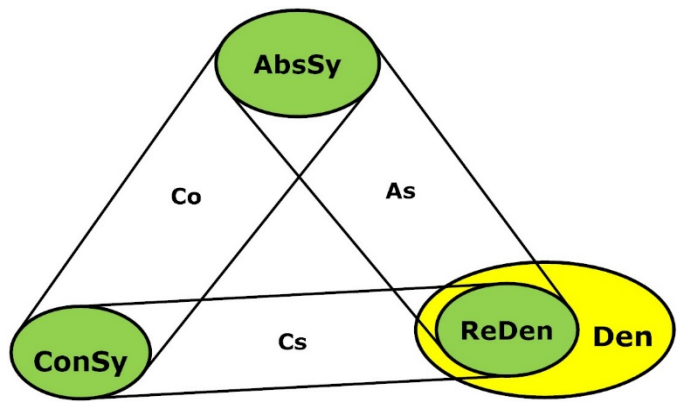

**Fig. 3.2-1 An algebraic model of a programming language**

As we saw in Sec.2.2, the abstract syntax is not very convenient for programmers. To make it more user-friendly, in the next step we build a concrete syntax ConSy. In typical situations, this is done by replacing prefix notation by infix notation and skipping some "unnecessary" parentheses. A typical example of skipping parentheses is the replacement of a sequential composition of instructions in the abstract-syntax:

   `;(ins-1, ;(ins-2, ins-3))`

by its concrete-syntax version:

   `ins-1 ; ins-2 ; ins-3`

Although the corresponding homomorphism Co (concretisation) is "gluing" two abstract programs

   `;(ins-1, ;(ins-2, ins-3))` and

```
;(;(ins-1, ins-2), ins-3)
```

into the same concrete program, this parsing ambiguity (of the corresponding grammar) is not harmful to the existence of a concrete semantics:

Cs : ConSy ↦ ReDen

since abstract semantics As is gluing these programs into a common denotation[10].

Another simplification that we may like to introduce into our language is the omission of parentheses in numeric expression. E.g. instead of writing

```
(x + (y + z)))
```

we would like to write

```
x + y + z
```
(3.2-1)

In this case, however, we end up with a syntax which does not have a semantics into Den, since the expression (3.2-1) corresponds to two concrete expressions:

```
(x + (y + z)))    and
((x + y) + z)
```

whose denotations are not the same. It is due to the fact that in every computer arithmetic there is a limit for the "size" of a number. E.g. if the largest acceptable number is 10, then

(-4 + (10 + 3)) = 'overload'     (an error-message, see Sec.2.4)

((-4 + 10) + 3) = 9

In other words, computer addition is not associative.

A usual solution in such a case is the assumption that expressions are evaluated from left to right which means that (3.2-1) is evaluated as

```
((x + y) + z).
```

In other words, an interpreter of the language first add the "missing" parentheses and then evaluates the expression according to the concrete semantics. The same technique is used in the evaluation of expressions with addition and multiplication, e.g.,

```
x + y + z * x
```

in which case the operation of adding parentheses refers to the priority of multiplication over addition, hence the resulting concrete expression is:

```
((x + y) + (z * x))
```

To formalize this technique in our framework we introduce yet another algebra called a *colloquial syntax* and denoted by ColSy (Fig. 3.2-2). This algebra is not homomorphic to concrete syntax and has a different signature. However, it is constructed in such a way there exists an implementable transformation

Rt : ColSy ↦ ConSy

---

[10] Formally this means that the algebra of concrete syntax is not more ambiguous than the algebra of denotation which guarantees the existence of a unique homomorphism between them (see Sec.2.13 of [16]).

which "removes colloquialisms", which in our case means adding the missing parentheses. Such a transformation is called the *restoring transformation* and of course, is not a homomorphism.

A user manual of a programming language with colloquialisms describes concrete syntax by a grammar, and the colloquialisms as additional grammatical clauses. This means that the programmer is free to use either a concrete syntax or a colloquial one.

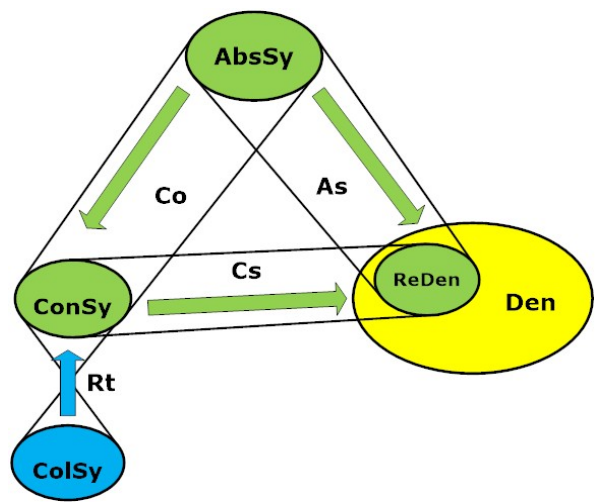

**Fig. 3.2-2 An algebraic model of a language with colloquial syntax**

To sum up, the construction of a programming language with a denotational model consists of five steps:

1. The construction of Den where we decide about the meaning of future programs and their constructors. This is the most creative step where we decide about all the programming mechanisms of our language.

2. The derivation of abstract syntax, i.e. its grammar, from the signature of Den. This step is fully programmable.

3. The definition of concrete syntax, i.e. its grammar. To a certain degree, this is a creative step again, although in this case it may be supported by a software tool which assists the designer in transforming the grammar of abstract syntax into its concrete counterpart.

4. The description of the semantics Cs of concrete syntax. The definition of this semantics, i.e. the semantic clauses as (2.2-1), may be derived algorithmically from the definitions of Den, AbsSy and ConSy.

5. The enrichment of the concrete syntax by colloquialisms and the definition of the corresponding restoring transformation. This is again a creative step.

## 3.3   Two layers of a programming language

In the sequel of the paper we will see how to use the described model to construct a programming languages with two basic layers of programming tools:

1. *applicative layer* covering data expressions and type expressions whose denotations are functions from states to data and from states to types respectively,

2. *imperative layer* covering instructions and declarations whose denotations are functions from states to states.

# 4  The applicative layer of Lingua

## 4.1  The data

Data available in **Lingua** may be split into two categories:

- *simple data* including Booleans, numbers, and words (finite strings of characters),
- *structural data* including list, many-dimensional arrays, records, and their arbitrary combinations.

Structural data may „carry" simple data as well as other structural data. That means that we may build "deep" data structures, e.g., lists that carry records of arrays. Lists and tables always carry elements of the same type whereas records are not restricted in this way.

All our data (with abstract errors) and the corresponding constructors constitute a many-sorted algebra of data.

Formally the data domains in **Lingua** are defined by the following set of so called *domain equations*:

| boo | : Boolean | = {tt, ff} |
| --- | --- | --- |
| num | : Number | — the set of all numbers with restricted decimal representations |
| ide | : Identifier | — a fixed finite subset of the domain Alphabet$^+$ |
| wor | : Word | = {'}Alphabet*{'} |
| lis | : List | = Data$^{c*}$ |
| arr | : Array | = Number $\Rightarrow$ Data |
| rec | : Record | = Identifier $\Rightarrow$ Data |
| dat | : Data | = Boolean \| Number \| Word \| List \| Array \| Record |

The symbols boo, num, ide etc. which precede our equations are metavariables that will run over the corresponding domains in further definitions. This is just another notational convention.

The domain Boolean consist of only two elements that represent "truth" and "false". The domains Alphabet, Number and Identifier, are the parameters of our model which means that they may differ from one implementation to another.

The Alphabet is a finite set of characters (except quotation marks), while Identifier is a finite fixed set of non-empty strings over Alphabet.

A word is a finite string (possibly empty) of the elements of Alphabet closed between apostrophes.

A list is a finite sequence (possibly empty) of arbitrary data.

An array is a mapping from numbers to data, and a record is a mapping from identifiers to data.

A data is a boolean, a number, a word, a list, an array or a record. Notice that identifiers are not included in data. They have been introduced only to define the domain of records. Identifiers that appear in records are called *record attributes*.

As we see, the four last equations have a recursive character, and therefore the existence of a solution of our set of equations is not evident. However, such a solution exists and is (in a sense) unique[11] which may be proved on the ground of the theory of chain-complete partially ordered sets (Sec. 2.7 of [16]).

It is to be emphasized in this place that the domain of data, and all of its subdomains, are larger than the corresponding sets of numbers, words, lists etc. that can be "generated" by the programs of **Lingua**. Further on we make sure that:

1. all "executable" data are restricted in their size — this is formalized be introducing a universal predicate oversized defined for all data,
2. for any given list or array all its elements are of the same type (see Sec.4.2),
3. the domain of each array must be of the form {1,…,n}, i.e. must be a set of consecutive positive integers starting from 1.

The constructors of data are defined in such a way that all reachable data satisfy the above restrictions. This technique allows keeping our domain equations relatively simple.

## 4.2     Composites, transfers, yokes, types and values

Every data in **Lingua** has a type. Types describe properties of data but represent entities which can be constructed and modified independently of data. Our mechanism of types allows programmers to define their own types for future use either in defining new types or in declaring variables[12].

Types are pairs consisting of a *body* and a *yoke*. Every type is associated with a set of data of that type called the *clan of the type*.

Intuitively a body describes an "internal structure of a data" — e.g., indicates that a data is a number, a list or a record — and formally is a combination of tuples and mappings. The domain equation that defines the domain of bodies is the following[13]:

bod : Body    =

    {('Boolean')} | {('number')} | {('word')} |                                 (*simple bodies*)

    {'L'} x Body |                                                                (*list bodies*)

    {'A'} x Body |                                                               (*array bodies*)

    {'R'} x (Identifier $\Rightarrow$ Body)                                      (*record bodies*)

The bodies of simple data are one-element tuples of metaconstants, e.g. ('Boolean'). The bodies of lists and arrays are respectively of the form ('L', bod) or ('A', bod) where the body bod is shared by all the elements of a list/array and where the *initials* 'L' and 'A' indicate that we are dealing with a list/ array.

A record body is of the form ('R', body-record) where body-record is a metarecord of bodies such as, e.g.:

    Ch-name ;        ('word'),

---

[11] It is unique in the sense that by the solution of such an equation we mean its least solution where the ordering is the componentwise set-theoretic inclusion .
[12] Technical details in Sec. 5.2 of [16].
[13] This is again a recursive equation (as it was the case of data-domain equations) and again its unique solution exists.

|  |  |
|---|---|
| fa-name ; | ('word'), |
| award-years ; | ('A', ('number')), |
| salary ; | ('number'), |
| bonus ; | ('number') |

The words on the left-hand-side of semicolons are attributes. The first two attributes and the last two have simple bodies, whereas the third one — an array body. For the sake of further discussion, the body defined above will be referred to as employee.

With every body bod, we associate a set of data with that body called *the clan of that body* and denoted by CLAN-Bo.bod. The function CLAN-Bo is defined inductively relative to the structure of bodies. E.g., the set CLAN-Bo.employee contains records with numbers, words, and one-dimensional number arrays assigned to the respective attributes.

Next important concept from the "world" of data and types is a *composite* which is a pair (dat, bod) consisting of a data and its body such that:

dat : CLAN-Bo.bod

Composites are the results of data-expression evaluations (Sec.4.4). The use of composites permits to describe the mechanism of checking if the arguments "delivered" to an operation are of appropriate types. E.g., if we try to put a word on a list of numbers, the corresponding operation will generate an error message.

Having defined composites, we can define *transfers* and *yokes*. Transfers are one-argument functions that transform composites or errors into composites or errors and *yokes* are transfers with Boolean composites as values. By a *Boolean composite* we mean (tt, ('Boolean')) or (ff, ('Boolean')). Yokes may also assume abstract errors as values.

Mathematically yoks are close to one-argument predicates on composites[14]. An example of a yoke that describes a property of composites whose bodies are employee may be the inequality:

**record.**salary + **record.**bonus < 10000,

This yoke is satisfied whenever its (unique) argument is a record composite with (at least) the attributes salary and bonus, and the data corresponding to these attributes satisfy the corresponding inequality. In this example

**record.**salary + **record.**bonus

is a transfer which is not a yoke. It transforms record composites into number composites. If the argument of this yoke/transfer is not a record with attributes salary and bonus that carry numbers, then the result of the computation is an error.

Yokes have been introduced into **Lingua** to describe SQL integrity constraints (for details see Sec.12 of [16]).

Transfers have merely a technical role. We need them only to define an algebra where yokes may be constructed. With every transfer we associate its clan:

CLAN-Tr.tra = (com | tra.com = (tt, ('Boolean'))}.

---

[14] They "are closed to predicates" rather than simply "are predicates" since they assume as values composites and abstract errors rather than just Boolean values tt and ff. Their logical constructors **and**, **or** and **not** are the three-valued constructors of John McCarthy's calculus defined by (Sec. 2.5).

Of course, the clans of transfers which are not yokes, are empty. By TT we denote the transfer that yields (tt, ('Boolean')) for any composite.

A pair that consists of a body and a yoke is called a *type*. For technical reasons, types are defined as pairs consisting of a body and an arbitrary transfer (i.e. not necessarily a yoke). With every type typ = (bod, tra) we associate its *clan* which is the set of such composites whose data belong to the clan of the body and which satisfy the transfer. Formally:

CLAN-Ty.(bod, tra) = {(dat, bod) | dat : CLAN-Bo.bod **and** (dat, bod) : CLAN-Tr.tra}

The last concept associated with data and types is *value*. A value is a pair (dat, typ), i.e. (dat, (bod, tra)), which we sometimes write as ((dat, bod), tra). As we see, a value may be regarded, either as a pair *data-type* or as a pair *composite-transfer*.

For technical reasons we also allow *pseudo-values* of the form (Ω, typ), where Ω is an abstract object called a *pseudo-data*.

Values are assigned in memory states to the identifiers of variables. Variable declarations assign pseudo values to variables, and initializing assignments replace Ω by a data.

As we are going to see, an assignment instruction — i.e., an instruction that assigns values to variables (see Sec.5.2) — may only change the data assigned to a variable, and in some special cases its body, but never its yoke. To change a yoke, we use special yoke-oriented instruction.

Summing up, the list of domains that are associated with data and their types in **Lingua** is the following

| | | |
|---|---|---|
| dat | : Data | = … (the definition in Sec.4.1) |
| bod | : Body | = … (the definition above in this section) |
| com | : Composite | = {(dat, bod) \| dat : CLAN-Bo.bod} |
| com | : BooComposite | = {(boo, ('Boolean')) \| boo : Boolean} |
| tra | : Transfer | = (Composite \| Error) ↦ (Composite \| Error) |
| yok | : Yoke | = (Composite \| Error) ↦ (BooComposite \| Error) |
| typ | : Type | = Body x Transfer |
| val | : Value | = Data x Type |

Similarly, as in many programming languages (although not in all of them), types in **Lingua** have been introduced for four reasons:

1. to define a type of a variable when it is declared, and to assure that this type remains unchanged (with some exceptions)[15] during program executions,

2. to ensure that a data which is assigned to a variable by an assignment is of the type consistent with the declared type of that variable,

3. to ensure that a similar consistency takes place when sending actual parameters to a procedure or when returning reference parameters by a procedure,

---

[15] These exceptions take place e.g. when we add a new attribute to a record or to a database table or if we remove such attribute.

4. to ensure that in evaluating an expression, an error message is generated whenever data "delivered" to that expression are of an inappropriate type, e.g., when we try to add a word to a number or to put a record to a list of arrays.

## 4.3  Expressions in general

Expressions are syntactic objects and their *denotations* are functions from states to composites (*data expressions*), to transfers (*transfer expressions*) or to types *type expressions*). In order to define these concepts we start with the definition of a *state*:

| | | | |
|---|---|---|---|
| sta | : State | = Env x Store | (state) |
| env | : Env | = TypEnv x ProEnv | (environment) |
| sto | : Store | = Valuation x (Error \| {'OK'}) | (store) |
| vat | : Valuation | = Identifier $\Rightarrow$ Value | (valuation)[16] |
| tye | : TypEnv | = Identifier $\Rightarrow$ Type | (type environment) |
| pre | : ProEnv | = Identifier $\Rightarrow$ Procedure \| Function | (procedure environment)[17] |

As we see, states are binding identifiers to values, to types, to procedures, or to functions (functional procedures) and besides they may store an error "in a dedicated register". If a state does not carry an error, then this register stores 'OK'. Every state is therefore a tuple of the form:

(env, (vat, err))     where err : Error | {'OK'}

Having defined states we can define the domains of expression denotations of three categories:

| | | |
|---|---|---|
| ded | : DatExpDen = State $\rightarrow$ Composite \| Error | (data-expressions denotations) |
| tra | : TraExpDen = Transfer | (transfer-expressions denotations) |
| ted | : TypExpDen = State $\mapsto$ Type \| Error | (type-expressions denotations) |

The denotations of data expressions are partial functions which is due to the fact that data expressions may include functional-procedure calls[18].

The fact that denotations of transfer expressions are just transfers rather than functions from states to transfers is a consequence of the fact that in our model transfers cannot be "stored" in states, as it is in the case for data and types. This is, of course, an engineering decision rather than a mathematical must. It has been assumed only for the sake of simplicity.

The three domains are the carriers of an *algebra of expression denotations* from which a syntactic (concrete) *algebra of expressions* is derived (as sketched in Sec.3.2) with the carriers DatExp, TraExp, TypExp. This leads to three functions of semantics which constitute a homomorphism between our two algebras.

Sde  : DatExp  $\mapsto$ DatExpDen

Stre : TraExp  $\mapsto$ TraExpDen

Ste  : TypExp  $\mapsto$ TypExpDen

---

[16] The metavariable running over valuations is "vat" since "val" has been reserved for values.
[17] The domains Procedure and Function are defined in Sec. 5.4
[18] Functional procedures may loop indefinitely and since this is not a computable property we cannot expect to have an error message in that case.

## 4.4 Data expressions

Data expressions evaluate to composites or errors. With every operation on data, we associate two constructors: of data-expression denotations and of data expressions. In this way, we define two mutually similar algebras and a homomorphism between them. This homomorphism is unique, is implicit in the definitions of both algebras and constitutes the semantics of data expressions. This section contains just one example of a syntactic constructor and of the corresponding semantic clause.

Consider the data operation of the numeric division **divide** and its syntactic counterpart "/". The clause of our grammar (Sec.2.3) that corresponds to the syntactic constructor is

(DatExp / DatExp)

In the sequel instead of dealing directly with grammatical clauses, I shall write them in the form of a *syntactic scheme*. In the present case:

(dae-1 / dae-2),

where `dae-1` and `dae-2` are metavariables denoting data expressions. The corresponding clause of the definition of semantics is shown below. The syntactic argument is closed in square brackets.

```
Sde.[(dae-1 / dae-2)].sta =
    let
        (env, (val, err)) = sta
    err ≠ 'OK'                  ➔ err
    Sde.[dae-i].sta = ?  ➔ ?                          for i = 1,2
    let
        num-i = Sde.[dae-i]. (env, (val, err))        for i = 1,2
    num-i : Error               ➔ num-i               for i = 1,2
    let
        (dat-i, bod-i) = num-i                        for i = 1,2
    bod-i ≠ ('number')          ➔ 'number-expected'   for i = 1,2
    dat-2 = 0                   ➔ 'division-by-zero'
    let
        dat-3 = divide(dat-1, dat-2)
    oversized.dat-3             ➔ 'overflow'
    true                        ➔ (dat-3, ('number'))
```

In the above definition the clause

Sde.[dae-i].sta = ? ➔ ? for i = 1,2

stands for

　　Sde.[`dae-1`].sta = ?

　　Sde.[`dae-2`].sta = ?

and analogously for all similar clauses. Intuitively our definition should be read as follows:

- If the input state carries an error, then this error becomes the final result of the computation.
- Otherwise, we evaluate both component expressions, and if one of these evaluations does not terminate, then (of course) the whole computation does not terminate.
- Otherwise, we check the bodies of both resulting composites and if one of them is not ('number'), then an appropriate error is generated.
- Otherwise, we check if the second argument of the division is zero, in which case an error is generated.
- Otherwise, we check if the result of the division is not oversized in which case an error is generated[19].
- Otherwise, the result of division becomes part of the resulting composite.

## 4.5　Transfer expressions

Transfer expressions evaluate to transfers or errors. Since transfers are not usual in programming languages — at least not as we define them — a few examples may be in order. Below the "current composite" means the composite which is the (only) argument of the transfer.

| | | |
|---|---|---|
| `273` | — | the resulting composite is (273, ('number')) independently of the current composite, |
| **record**.`price` | — | if the current composite carries a record with an attribute `price`, its body ('number') and its data dat, then the resulting composite is (dat, ('number')), and otherwise is an error. |
| **all-list** `number` **ee** | — | this is a yoke; if the current composite does not carry a list, then an error is generated, otherwise, if it is a list of numbers then the resulting composite is (tt, ('Boolean')), and otherwise, it is (ff, ('Boolean')), |
| **record**.`price + `<br>　**record**.`vat < 1000` | — | this is a yoke; if the current composite does not carry an appropriate record, then error and otherwise, if the sum of data assigned to `price` and `vat` is less than 1000, then (tt, ('Boolean')), and otherwise (ff, ('Boolean)) |

Now let us consider a transfer expression with the asyntactic scheme

　　**all-list** `tre` **ee**.

---

[19] In our definitions this part of procedure is described in an abstract way, but the implementation does not need to preform it literally, i.e. by first dividing the given numbers and only then checkig, if that was possible. In an implementation a programmable solution should be chosen.

Such an expressions is satisfied if all elements of a current list satisfy the transfer `tre`. The semantic clause is the following:

Stre.[**all-list** `tre` **ee**].com =

    com : Error → com

    sort.com ≠ 'L' → 'list-expected'

        **let**

            ((dat-1,…,dat-n), ('L', bod)) = com    (list elements always have the same body)

            com-i = Stre.[`tre`].(dat-i, bod)        for i = 1;n

        com-i : Error → com-i    for i = 1;n

        **not** com-i : BooComposite → 'a-yoke-expected'

        ($\forall$ i = 1;n) com-i = (tt, ('Boolean')) → (tt, ('Boolean'))

        **true** → (ff, ('Boolean'))

This definition may be intuitively read as follows:

1. If the current composite is an error, then the result is this error.
2. Otherwise, if the current composite does not carry a list, then an error is signalized.
3. Otherwise, the transfer Stre.[tre] is applied to composites created from the data dat-i of the list and the "internal body" bod of the list. Notice that lists carry data, rather than composites.
4. If one of these composites is an error, then the first such an error is the result of the computation.
5. If one of these composites is not a Boolean composite, then an error is generated.
6. If all resulting composites are (tt, ('Boolean')), then the resulting composite is (tt, ('Boolean')), and otherwise, it is (ff, ('Boolean')).

## 4.6 Type expressions

Type expressions evaluate to types or errors. E.g., the denotation of the type expression:

```
record-type
    Ch-name       as word,
    fa-name       as word,
    birth-year    as number,
    award-years   as number-array,
    salary        as number,
    bonus         as number
  ee
```

is a function on states that creates a record type or generates an error. This expression refers to two built-in types `word` and `number` and one user-defined type `number-array` (arrays of numbers).

Now consider an example of a syntactic scheme of an expression that creates a one-attribute record type:

    **record-type** ide **as** tex **ee**

where ide is an identifier and tex is a type expression. The corresponding semantic clause is the following:

    Ste.[ **record-type** ide **as** tex **ee** ].sta =

      **let**

        (env, (val, err)) = sta

      err ≠ 'OK' ➔ err

      **let**

        typ = Ste.[tex]. sta

      typ : Error ➔ num-i

      **true**       ➔ (('R', [ide/typ]), TT)

This clause is read as follows:

1. If the input state carries an error, then this error becomes the result of the computation.
2. Otherwise, we compute the type defined by tex, and if it is an error, then this error becomes the result of the computation.
3. Otherwise, the resulting type is the record type (('R', [ide/typ]), TT).

To construct a many-attribute record type we use the operation of adding an attribute to a given record type with the following syntactic scheme:

    **expand-record-type** tex-1 **at** ide **by** tex-2 **ee**

and to replace a current transfer of an arbitrary type defined by tex, by a new transfer tre, we use a type expression with a scheme:

    **replace-transfer-in** tex by tre **ee**

## 4.7   The concrete syntax of expressions

The full grammar of the syntax of expressions in shown in Sec.5.4.2 of [16][16]. Below only an excerpt of it:

    dae : DatExp =

      true | false | number | word |

      Identifier | (DatExp **and** DatExp) | (DatExp **or** DatExp) …|

      (DatExp + DatExp) | (DatExp / DatExp) | DatExp **glue** DatExp |

      **list** DatExp **ee** | **push** DatExp **on** DatExp **ee** | **top** (DatExp) |

      …

      **if** DatExp **then** DatExp **else** DatExp **fi**

In the first line of this clause, the metavariables number and word represent the fact that all numbers and words up to a certain size are acceptable as expressions. At the level of implementation, an appropriate lexical analyser is defined. The keyword **`glue`** corresponds to the concatenation of words.

    tre : TraExp =

        num | wor | (TraExp + TraExp) | (TraExp / TraExp) |

        **sum** (TraExp) | **max** (TraExp) |

        …

    tex :TypExp =

        `boolean` | `number` | `word` |

        Identifier | **`list-type`** TypExp **`ee`** | **`array-type`** TypExp **`ee`** |

        **`record-type`** Identifier **`as`** TypExp **`ee`** |

        …

In the syntax of type expressions `number` and `word` denote themselves, i.e. the names of simple types.

## 4.8  The colloquial syntax of expressions

As was already explained, colloquial syntax includes all concrete syntax which means that the use of colloquialisms is optional. On the algebraic level, each colloquialism is a new constructor, which makes the algebra of colloquial syntax not similar to the algebra of concrete syntax. Below three examples of colloquialisms described informally:

1. `x or y or z` means `(x or (y or z))`,
2. `x + y + z + x*y` means `(x + y) + z) + (x*z)`
3. **`array`** `[x, x+y, 3*y]` means

    **`add-to-arr`**

        **`add-to-arr`**

            **`array`** `x` **`ee`**

        **`new`** `x+y` **`ee`**

    **`new`** `3*y` **`ee`**

# 5  The imperative layer of the language

Expressions of all types belong to an applicative layer of **Lingua**. Their denotations use states as arguments but neither create them nor change. The latter tasks are performed by *instructions, variable declaration, procedure- and function declarations* and by *type definitions*. All of them belong to an imperative layer of the language.

## 5.1  Some auxiliary concepts

Two new metapredicates are necessary to define the semantics of the imperative layer of our language.

The metapredicate

    is-error : State $\mapsto$ {tt, ff}

returns tt whenever a state carries an error.

We say that body bod-1 *is coherent* with bod-2, in symbols

    bod-1 <u>coherent</u> bod-2

whenever:

1. bod-1 = bod-2 or
2. they are record bodies, and one of them results from the other by adding or by removing an attribute.

We also introduce an operator of inserting an error into a state:

    ◄ : State $\mapsto$ State

    (env, (vat, err)) ◄ error = (env, (vat, error))

## 5.2 Instructions

Instructions change states, and therefore instruction denotations are partial functions from states to states:

    ind : InsDen = State $\to$ State

The partiality results from the fact that the execution of an instruction may be infinite (an instruction may loop). The semantics of instructions is a function

    Sin : Instruction $\mapsto$ InsDen

Contrary to expression denotations which may generate an error, instruction denotations write an error into the error register of a state. The denotations of the majority of instructions are *transparent* relative to error-carrying states, i.e., they do not change such a state but only pass it to the subsequent parts of the program. However, an error may also cause an error-handling action (see Sec.6.1.8 of [16][16]).

The basic instruction is, of course, an *assignment* of a value to a variable identifier. The syntactic scheme of an assignment is:

    `ide := dae`

and the corresponding semantic clause is the following:

    Sin.[`ide := dae`].sta =

        is-error.sta                        ➔ sta

        **let**

            ((tye, pre), (vat, 'OK')) = sta

        `vat.ide` = ?                 ➔ sta ◄ 'identifier-not-declared'

        Sde.[`dae`].sta = ?          ➔ ?                       (an infinite execution)

        Sde.[`dae`].sta : Error    ➔ sta ◄ Sde.[`dae`].sta

        **let**

    ((dat-f, bod-f), tra) = vat.`ide`                                                    (f – former)

    (dat-n, bod-n)      = Sde.[`dae`].sta                                             (n – new)

    com                    = tra.(dat-n, bod-n)

com : Error                      ➔ sta ◄ com

**not** bod-n coherent bod-f     ➔ sta ◄ 'no-coherence'

**not** com : BooComposite     ➔ sta ◄ 'a-yoke-expected'

com = (ff, ('Boolean')         ➔ sta ◄ 'yoke-not-satisfied'

**let**

    val-n = ((dat-n, bod-n), tra)

**true**                           ➔ ((tye, pre), (vat[ide/val-n], 'OK'))

The denotation of an assignment changes an input state into an output state in nine steps:

1. If an input state carries an error, then this state becomes the output state.
2. Otherwise, if the identifier `ide` has not been declared, i.e., if no value or a pseudo value has been assigned to it in the valuation val, then an error message is loaded to the error register.
3. Otherwise, if an attempt to evaluate the data expression leads to an infinite execution, then (of course) the executions of the instruction is infinite as well.
4. Otherwise, if the expression evaluates to an error, then this error is loaded to the error register of the state.
5. Otherwise, it the transit applied to the new composite returns an error, then this error is loaded to the error register.
6. Otherwise, if the composite computed from the expression has a body non-coherent with the body of the identifier's type, then an error is loaded to the error register.
7. Otherwise, if the composite computed by the transit is not Boolean, i.e. if the transit was not a yoke, then an error is loaded to the error register.
8. Otherwise, if the yoke is not satisfied, then an error message is loaded to the error register.
9. Otherwise, the new value is the new composite and the current (i.e. not changed) yoke, and this new value is assigned to the identifier `ide`.

Notice that as a consequence of the claim 6. together with the definition of the coherence of bodies (Sec.5.1) an assignment may change the body of a value assigned to a variable only if this body is a record, and only by adding or by removing an attribute to/from that record.

The remaining instructions belong to one of the following seven categories where the first four are *atomic instructions,* and the other three are *structural instructions*, i.e., instructions composed of other instructions and expressions:

1. the replacement of a yoke assigned to a variable by another one
   **yoke** `ide := tre`,

2. the empty instruction
   **skip**,

3. the call of an imperative procedure
   **call** ide (**ref** apar-r **val** apar-v)
   where apar-r and apar-v are lists (maybe empty) of identifiers called respectively *actual reference-parameters* and *actual value-parameters*,

4. the activation of an error-handling
   **if** dae **then** ins **fi**,

5. the conditional composition of instructions
   **if** dae **then** ins-1 **else** ins-2 **fi**,

6. the loop
   **while** dae **do** ins **od**,

7. the sequence of instructions
   ins-1 ; ins-2.

In the yoke-replacement instruction, the new value of the identifier ide gets the old composite but a new transfer. This transfer must be satisfied with the current composite[20].

The empty instruction **skip** is needed to make functional-procedure declarations sufficiently universal; this will be seen in Sec.5.4.

The discussion of procedures is postponed to Sec.5.4

The error handling is activated if the current state carries an error, i.e. a word, that is equal to the word that the data-expression dae evaluates to. If this happens, the "internal" instruction ins is executed for a state that results from the initial state where the current error has been replaced by 'OK'[21].

The semantics of the three remaining categories of instruction is usual with the exception that in the last two cases an expression may generate an error message. In such a case that error is stored in the error register of the state.

## 5.3    Variable declarations and type definitions

*Variable-declaration denotations* are total functions that map states into states:

vdd : VarDecDen = State $\mapsto$ State

assigning types to identifiers and leaving their data undefined, i.e. assigning pairs of the form (Ω, typ). The syntactic scheme of a single declaration is of the form:

**let** ide **be** tex **tel**

Variable declarations are similar to assignments with the difference that for a declaration an error 'identifier-not-free' is signalized whenever the identifier ide is bound in the input state. It means that a variable may be declared in a program only once. During program execution the value assigned to a variable may be changed only by changing:

- the composite of the value by an assignment instruction,

---

[20] This instruction has been introduced mainly for the sake of SQL tables discussed in [16].
[21] For details see Sec.6.1.8 of [16].

- the yoke of the value by a yoke-replacement.

Type definitions are of the form

    **set** ide **as** tex **tes**

and their denotations are similar to those of variable declarations, i.e.

    tdd : TypDefDen = State $\longmapsto$ State

with the difference that instead of assigning a pseudovalue to a variable identifier in a valuation they assign a type to a type-constant identifier in a type environment.

An identifier that is bound to a type in a state is called a *type constant*. Notice that "a constant" rather than "a variable" since a type once assigned to an identifier, cannot be changed in the future (an engineering decision).

Similarly as in the case of assignments, also type definitions, and variable declarations may be combined sequentially using a semicolon constructor.

## 5.4 Procedures

Procedures in **Lingua** may be *imperative* or *functional*. The former are functions that take two lists of actual parameters — *value parameters* and *reference parameters* — and return partial functions on stores[22]. Functional procedures take only value parameters and return partial functions from states to composites or errors:

    ipr : ImpPro = ActPar x ActPar $\longmapsto$ Store $\rightarrow$ Store

    fpr : FunPro = ActPar $\longmapsto$ State $\rightarrow$ (Composite | Error)

In these equations, ActPar is a domain of *actual-parameter lists* defined by the domain equation:

    apa : ActPar = () | Identifier | ActPar x ActPar

As we see, actual-parameter lists are finite (maybe empty) sequences of identifiers. In turn, formal-parameter lists that appear in procedure declarations are finite (maybe empty) sequences of pairs consisting of an identifier and a type-expression denotations:

    fpa : ForPar = () | Identifier x TypExpDen | ForPar x ForPar

Returning to procedures, notice that we do not talk here about "procedure denotations" but about "procedures" as such since they are purely denotational concepts. In other words, they do not have syntactic counterparts. At the level of syntax, we have only *procedure declarations* and *procedure calls* which, of course, have their denotations.

A syntactic scheme of an imperative-procedure declaration is of the following form (the carriage returns are of course syntactically irrelevant):

    **proc** ide (**ref** fpar-r **val** fpar-v)
      pro
    **end proc**

---

[22] The fact that procedures transform stores rather than states is a technique (introduced in [17]) that allows to define recursion in avoiding the selfapplication of procedures, i.e. a situations where a procedure takes itself as an argument. Of course, procedure calls are instructions and therefore they transform states into states.

where `pro` is a program (see later) and `fpar-r` and `fpar-v` are the lists of respectively formal reference-parameters and formal value-parameters. A syntactic example of a list of formal parameters may be as follows:

  (**val** age, weight **as** number, name **as** word,

    **ref** patient **as** patient-record)

Expressions different from single-identifier-expressions are not allowed as value parameters since such a solution would complicate the model as well as program-construction rules (an engineering decision).

If we want to declare a group of mutually recursive procedures, we use a *multiprocedure declaration* of the form:

  **begin multiproc**

    ipd-1;

    ipd-2;

    …

    ipd-n

  **end multiproc**

where the `ipd`'s are imperative-procedure declarations. Intuitively this means that these procedure declarations have to be elaborated (compiled) "as a whole", rather than one after another (details in Sec.7.4 of [16]).

The syntactic scheme of a functional-procedure declaration is of the form :

  **fun** ide (fpar)

    pro

  **return** dae **as** tex

A call of a functional procedure declared in this way first executes the program `pro` and then evaluates the data expression `dae` in the output state of the program. If the composite generated by that expression is of the type defined by the type expression `tex`, then this composite becomes the result of the call of the function. Otherwise, an error is signalled.

In particular, the program in a functional-procedure declaration may be the trivial instruction **skip** — which "does nothing" — and the exporting expression may be a single identifier.

The (concrete) syntactic schemes of an imperative-procedure call and a functional-procedure call are respectively:

  **call** ide (**ref** apar-r **val** apar-v)    — imperative-procedure call

  ide (apar-v)    — functional-procedure call

Notice that the second call has no reference parameters since functional procedures do not have any side-effects — they do not modify a state (an engineering decision).

All types and procedures defined in the hosting program <u>before</u> (see Sec.5.4) the declaration of a procedure are visible in the body of this procedure, and therefore they do not need to be passed as parameters (an engineering decision).

In the version of **Lingua** described in the present paper procedures cannot take other procedures as parameters. However, it is shown in Sec. 7.6 of [16] how to construct a hierarchy of procedures that can take procedures of lower rank as parameters. This construction protects procedures from taking themselves as parameters which would lead to non-denotational models (a mathematical decision).

## 5.5 The execution of a procedure call

In the descriptions of procedure mechanisms, we use some concepts having to do with the fact that procedures are created when they are declared and are executed when they are called. In respect to that, we shall talk about states (and their components) of a *declaration-time* and of a *call-time* respectively[23]. Traditionally by a *procedure body,* we mean the program that is executed when a procedure is called.

As has been already announced, there are no global variables in procedures (an engineering decision)[24]. The intention is that the head of a procedure-call describes explicitly and completely the communication mechanisms between a procedure and the hosting program. That solution may seem restrictive but — in my opinion — guarantees a better understanding of program functionality by programmers and definitely simplifies program-construction rulers.

Execution of a procedure call may be intuitively split into four stages illustrated in Fig. 5.5-1. (formal definitions in Sec.7.3 of [16]).

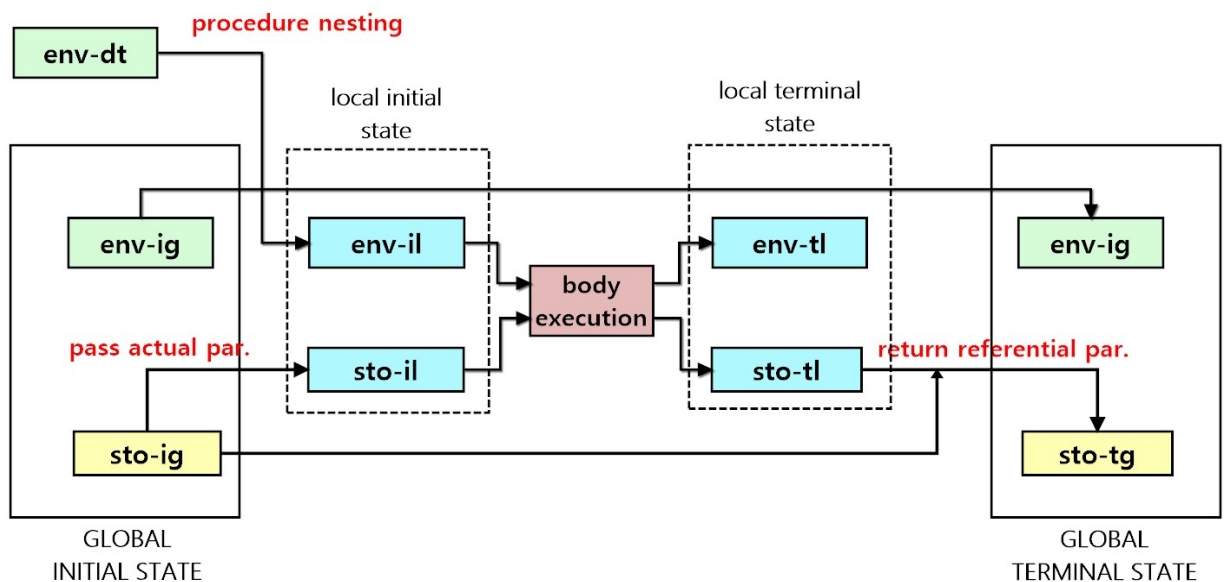

**Fig. 5.5-1 The execution of a procedure call**

1. **The inspection of an *initial global state* —** that state consists of:

    a. an *initial global environment* env-ig,

    b. an *initial global store* sto-ig = (vat-ig, err)

    If err ≠ 'OK', then the initial global state is returned by procedure call and therefore becomes the terminal global state. In the opposite case, an initial local state is created.

---

[23] These ideas, similarly to a few others, have been borrowed from M. Gordon [22]
[24] If we would like to introduced global variables, we should define the local store of a procedure call as a modification of its global store.

2. **The creation of an *initial local state*** — that state consists of:

    a. *initial local environment* env-il created from the <u>declaration-time environment</u> by nesting in it the called procedure; this nesting is necessary to enable recursive calls,

    b. *initial local valuation* vat-il covering only formal parameters with assigned values of corresponding actual parameters; to get the latter values, we refer to initial global valuation val-ig.

3. **The transformation of the local initial state** by executing the procedure body. If this execution terminates, then the local terminal state consists of:

    a. *terminal local environment* env-tl,

    b. *terminal local store* sto-tl = (val-tl, err-tl).

    If err-tl ≠ 'OK', then a global terminal state is created from the initial global-state by loading to it err-tl. Notice that in this case, the terminal local-environment and the terminal local store are "abandoned". Otherwise, the terminal global state is created.

4. **The creation of the *terminal global state*** — that state consists of:

    a. *initial global environment* env-ig; notice that terminal local environment env-tl is "abandoned",

    b. *terminal global store* sto-tg created from initial global store sto-ig by "returning" to it the values of formal referential parameters (stored in sto-tl) and assigning them to the corresponding actual referential parameters.

Notice that initial local environment "inherits" all types and procedures from the declaration-time environment. Procedure body may use its own local environment types and procedures, but after the completion of the call, they cease to exist, since the hosting program returns to the initial global environment.

It is to be underlined that the procedure body may access only that part of the environment which was created <u>before the procedure declaration</u>.

Of a similar character is the local valuation that is created only in procedure execution-time, although in this case the values or reference-parameters stored in it are eventually returned to the terminal global valuation.

Summarizing the visibility rules concerning procedure call:

1. the only variables visible in procedure-body are formal parameters plus variables local to the body (declared in it),

2. the only types and procedures visible in procedure-body are declaration-time types and procedures plus locally declared ones,

3. variables, types and procedures declared in the procedure-body are not visible outside of procedure call.

All these choices are not mathematical necessities but pragmatic engineering decisions dictated by the intention of making our model relatively simple which should contribute to the simplicity of program-construction rules and to a better understanding of programs by language-users.

Procedures in **Lingua** may call themselves recursively either directly or indirectly. At the level of semantic clauses, this leads to recursive definitions of the denotations of procedure declarations. For formal definitions see Sec.7.3.2 in [16][16].

## 5.6    Preambles and programs

Each program in **Lingua** consists of a preamble followed by an instruction. The syntactic scheme of a program is therefore of the form:

   **begin-program** pam ; ins **end-program**

where pam is a preamble.

Preambles are sequential compositions of type-constant definitions, data-variable declarations and procedure declarations. Their syntax is defined by the following grammatical clause:

pam : Preamble =

   ImpProDec | MultiProDec | FunProDec | TypDef | VarDec | **skip** |

   Preamble ; Preamble

Similarly to instructions also preambles contain **skip** which represent an identity state-to-state function. The semantics of programs and preambles are the following functions:

   Spr   : Program    $\mapsto$ ProDen

   Spre : Preamble   $\mapsto$ PreDen

which are defined by structural induction:

   Spr.[pam ; ins] = Spre.[pam] • Sin.[ins]

and

   Spre.[ipd]            = Sipd.[ipd]

   Spre.[mpd]           = Smpd.[mpd]

   …

   Spre.[pam-1 ; pam-2]   = Spre.[pam-1] • Spre.[pam-2]

Intuitively the clauses for preambles are read as follows:

- the semantics of preambles applied to imperative-procedure declarations coincide with the semantics of such declarations,
- the semantics of preambles applied to multi-procedure declarations coincide with the semantics of such declarations,
- …
- the denotation of a sequential composition of preambles is a sequential composition of their denotations.

Programs with the trivial preamble **skip** — if executed "without a context" — will always generate an error, unless they (the programs) are the **skip** themselves. Such programs are allowed because they may appear in procedure declarations as the bodies of procedures without locally declared objects. In turn, programs with trivial preambles and instructions are allowed in the declarations of functional procedures[25].

---

[25] Both these solutions, although in a slightly different form, have been suggested to me by Andrzej Tarlecki.

## 5.7 The carriers of our algebra of denotations

These carriers are listed below. For each of them there is a corresponding carrier in the algebra of syntax.

| | | | |
|---|---|---|---|
| ide | : Identifier | | (identifiers) |
| ded | : DatExpDen | = State → CompositeE | (data-expression denotations) |
| tra | : TraExpDen | = Transfer | (transfer-expression denotations) |
| ted | : TypExpDen | = State ↦ TypeE | (type-expression denotations) |
| vdd | : VarDecDen | = State ↦ State | (variable-declaration denotations) |
| tdd | : TypDefDen | = State ↦ State | (type-constant denotations) |
| ind | : InsDen | = State → State | (instruction denotations) |
| fpa | : ForPar | = (Identifier x TypExpDen)$^{c*}$ | (formal parameters) |
| apa | : ActPar | = Identifier$^{c*}$ | (actual parameters) |
| ipc | : IprComponents | = Identifier x ForPar x ForPar x ProDen | (imperative-procedure components) |
| cmp | : MprComponents | = IprComponents$^{c+}$ | (multiprocedure components) |
| ffc | : FprComponents | = Identifier x ForPar x ProDen x DatExpDen x TypExpDen | (functional procedure components) |
| idd | : IprDecDen | = State ↦ State | (imperative-procedure-declarations denotations) |
| mpd | : MulProDecDen | = State ↦ State | (multiprocedure-declarations denotations) |
| fdd | : FprDecDen | = State ↦ State | (function-declaration denotations) |
| pde | : PreDen | = State → State | (preamble denotations) |
| prd | : ProDen | = State → State | (program denotations) |

# 6 Lingua-SQL

## 6.1 General assumptions about the model

The denotational model of **Lingua-SQL** is built as an extension of the model of **Lingua** by adding:

1. new data domains corresponding of databases, tables, rows, and specific SQL-data,
2. new constructors defined on these domains.

## 6.2 Data, bodies and composites

So far values in **Lingua** consisted of a composite and a transfer. This principle is kept in **Lingua-SQL** for values carrying simple data, rows and tables but in the case of databases, values are records of tables supplemented by graphs of subordination relations (Sec. 6.6).

In **Lingua-SQL** lists, records and arrays do not carry rows, tables and databases and table fields do not contain lists, records and arrays. On the other hand, the extended repertoire of simple SQL values is available for the constructors of lists, records and arrays.

Simple data which are new in **Lingua-SQL** are associated with time, i.e. with calendars and clocks:

dat  : Date      = Year x Month x Day

tim  : Time      = Hour x Minute x Second

dti  : DateTime  = Date x Time

where Year, Month, Day, Hour, Minute and Second are defined as finite sets of numbers in an obvious way. Since simple data play a special role in SQL, we need a domain of such data:

sda : SimData = Boolean | Number | Word | Date | Time | DateTime | {Ө}

All former constructors with simple data as arguments — e.g. that add a new attribute to a record — are extended in an obvious way to the new domain.

To include rows and tables with empty fields in our model, we introduce an *empty data* Ө[26]. This data will never appear as a value of an expression and will never be assigned to a variable.

With the extended set of simple data, we can extend the set of corresponding operations, e.g. by allowing to add a number to a date. I do not define such operations explicitly assuming that their class is a parameter of our model.

The subcategories of numbers such as INTEGER, SMALLINT, BIGINT, DECIMAL(p, s), or of words CHARACTER(n), CHARACTER VARYING(n), BLOB, will correspond to yokes rather than to types.

As was already announced we introduce two new sorts of structural data:

row : Row    = Identifier $\Longrightarrow$ SimData

tab : Table  = Row$^{c*}$

At the level of domain equations, tables may contain rows of different length and different attributes. However, such tables will not be reachable in the algebra of composites.

Data bases do not appear at the level of data. They are defined only at the level of values (Sec.6.6).

Similarly, as in **Lingua**, all SQL data have corresponding bodies. The bodies of new simple data are defined as one-element tuples of words, hence:

sbo : SimBody = {('Boolean'), ("number"), ('word'), ('date'), ('time'), ('date-time')}

The bodies of new structural data are defined by the equations:

bod : RowBody = {'Rq'} x RowRec

ror  : RowRec  = Identifier $\Longrightarrow$ SimBody

bod : TabBody = {'Tq'} x Row x RowBody

As one can guess from these definitions, the composites of rows in a table will have a common body. The row contained in a table body carries the information about default data for columns.

---

[26] Notice that Ө, which is assignable to fields of rows and tables, is different from Ω which is assigned to a variable at the declaration-time.

Its list of attributes must coincide with the list of the attributes of the corresponding row body. This property will be insured by table-body constructors.

The domain **BodyE** is extended by new simple bodies and the bodies of rows and tables.

The function **CLAN-Bo** from **Lingua** is extended in an obvious way on row bodies. In the case of table bodies, we assume that each row of a table must have an appropriate record structure and that in each field with a non-empty default value there is a non-empty value. Of course, it does not need to be a default value. The latter are used when adding to a table a new row or a new column.

We assume that the empty table — a table with an empty tuple of rows — belongs to the clan of every table body.

The domain **CompositeE** is appropriately extended by composites associated with new simple data, row data, and table data. Additionally, we introduce an auxiliary domain of simple composites:

com : SimCom = {(dat, bod) | (dat, bod) : CompositeE **and** bod : SimBody}

and we also assume that (Ө, bod) is a composite for every simple bod.

## 6.3   The subordination of tables

Subordination relations describe the binary relationships that can hold between tables. Let then A and B be tables and let ide be an attribute that appears in both of them. Let A.ide and B.ide be the corresponding columns in these tables.

We say that A *is subordinate to* B *at* ide or that A is a *child* and B is a *parent,* that we write as

A sub[ide] B

if the following three conditions are satisfied:

1. an ide-column appears in both tables; the identifier ide is called the *subordination indicator*,
2. the column B.ide is repetition-free,
3. the column A.ide contains only the data that appear in B.ide

The points 2. and 3. together mean that each row of A unambiguously points to a row in B. By a *subordination graph* we mean any finite set of triples of identifiers:

sgr : SubGra = Sub.(Identifier x Identifier x Identifier)[27]

Each tuple (ide-c, ide, ide-p) in sgr is called an *edge of the subordination graph*, where ide-c (child) and ide-p (parent) play the role of graph nodes, and ide is a label of the edge. In the context of a given state, each edge expresses the fact that a subordination relation holds between the tables named ide-c and ide-p where ide is the subordination indicator.

About the subordination graphs, we assume only that ide-c ≠ ide-p, although such graphs may contain cycles. Notice also that there may be many edges starting in one node (one child may have many parents), and many edges may end in one node (many children may have a common parent).

---

[27] Notice that since the set Identifier is finite, each subordination graph is finite as well.

## 6.4    Transfers

Types — as we understand them in this paper — are mentioned in SQL-manuals only in the context of simple data and even in that case in a very unclear and incomplete way. The types of tables are implicit in table declarations, and the types of rows, columns and databases are totally absent. In table declarations, the descriptions of bodies are mixed with the description of yokes, and with database instructions, and are called *integrity constraints*.

Unfortunately, in none of the known to me SQL manuals (their list is given in the preamble to Sec.11 of [16][16]), I have found a complete description of integrity constraints. Although all of them have a certain common part, besides that part, each manual offers different ideas. In this situation, I decided to construct such a model of SQL types that would cover a "sufficiently large" spectrum of types that appear in SQL applications.

Since in **Lingua-SQL** there are no database composites, there will not be database transfers either. The properties of databases will be described by:

- the yokes referring to their tables,
- subordination graphs which are only seen at the level of values.

We assume that in **Lingua-SQL** we have all so-far-defined transfer-constructors, and in particular — Boolean constructors. New constructors will generate transfers on new simple composites — these are regard as the parameters of our model — plus row- and table-transfers.

The row transfers are analogous to record transfers of **Lingua**. Table transfers split into two classes.

The first contains *quantified table-yokes* which describe table properties by row yokes that should be satisfied for all rows of a table.

Notice that although quantified table-yokes express properties of table-rows explicitly, they express implicitly — due to quantifiers — some properties of columns, such as, e.g., that each element of a column is a number. This technique does not allow, however, to express properties of columns regarded as a whole, e.g. that a column is ordered or that it does not contain repetitions. To express such properties, we need special column-dedicated yoke constructors. Here is one example of such a constructor:

no-repetitions-tb : Identifier ⟼ Transfer

no-repetitions-tb.ide.com =

    com : Error          ➔ com

    sort.com ≠ 'Tq'    ➔ 'table-expected'

    **let**

       col = Cc[get-co-from-tb].(ide, com)

    col : Error           ➔ col

    **true**                   ➔ no-repetitions.col

We create a tuple of composites col which represents the column of the attribute ide, and then we check if this tuple satisfies a universal predicate no-repetitions. The created column does not contain the element that corresponds to the row of default values.

Since we have Boolean constructors among the constructors of yokes, we can use them to construct yokes that express properties of several columns of a table and all of its rows. Notice that contrary to the SQL standard the properties of columns and rows may be combined by arbitrary Boolean constructor rather than by conjunction only[28].

## 6.5    Types

The algebra of types of **Lingua-SQL** contains four carriers:

- Identifier
- Transfer
- CompositeE
- TypeE

and besides the constructors already defined for **Lingua** contain three groups of new constructors:

1. new transfer constructors (Sec.6.4),
2. selected constructors of row composites needed to construct the rows of default values,
3. three type constructors: of creating a one-attribute row, of adding an attribute to a row and of creating table type.

Row types are created similarly as record types with the difference that now the added type must be simple.

## 6.6    Database values

Database values are defined as pairs consisting of an (intuitively understood) record of table values and a subordination graph (Sec.6.3). About databases we assume additionally the following:

- to make a database accessible in a program, its tables must be assigned to variable identifiers in the current valuation,
- in every state its valuation carries tables of only one database; this database is called the *active database*.

To describe this mechanism new notions are necessary.

According to our assumptions we expand the current domain of simple values and we introduce the domains of row values and table values:

RowVal = {(com, tra) | sort.com = 'Rq' **and** tra.com = (tt, ('Boolean'))}

TabVal = {(com, tra) | sort.com = 'Tq' **and** tra.com = (tt, ('Boolean'))}

By a *database record* we mean a mapping that maps identifiers into table values:

dbr : DatBasRec = Identifier $\Longrightarrow$ TabVal

Of course, database records are not records in the sense of Sec.4.1, but only in a set-theoretic sense.

---

[28] To say the truth I am not sure if such a generalisation has a practical value.

We say that a database record dbr *satisfies the subordination relation* identified by a subordination graph sgr, in symbols

   dbr <u>satisfies</u> sgr,

if for every edge (ide-c, ide, ide-p) of the graph, the tables assigned to ide-c and ide-p are defined, i.e.

   (com-c, tra-c) = dbr.ide-c

   (com-p, tra-p) = dbr.ide-p

and the subordination relation holds, i.e.

   com-c sub[ide] com-p

By a *database value* we mean a pair consisting of a database record and a subordination graph that describes the subordination relations satisfied by that record:

   dbv : DbaVal = {(dbr, sgr) | dbr <u>satisfies</u> sgr}

We may say that for database values, the role of a yoke is played by the predicate <u>satisfies</u>. Notice, however, that since a database record caries table values, the tables of the database satisfy their own yokes.

## 6.7   States

Similarly as in **Lingua,** states in **Lingua-SQL** bind values with variables and types with type constants. The general definitions of types and values remain as in Sec.4.2 except for database values (Sec. 6.6). Consequently, the values in **Lingua-SQL**, i.e., objects which may be assigned to variable identifiers are all the values of **Lingua**, and additionally the values that carry:

1. simple SQL data,
2. rows,
3. tables,
4. databases.

Of course, database values are not values in the former sense of the word since they are not composed of a data and a type. The type of a database is implicit in the types of its tables and in the subordination graph.

In every state several data bases may be stored, i.e. assigned to identifiers, but only one base may be active at a time, i.e. the tables of only one base may be assigned to identifiers in valuations.

For states I assume the existence of four system identifiers:

   sb-graph   — that binds the subordination graph of the active base in the environment,

   copies     — that binds a finite sets of table names (identifiers) in the valuation,

   monitor    — that binds one table in the valuations, (the table displayed on a monitor)

   check      — that binds words 'yes' and 'no' in valuations.

Their role will be explained later. So far we assume only that they cannot be used as identifiers of variables, of type constants and of procedures. The identifier check is called the *security flag*.

The signature of the algebra of denotations of **Lingua-SQL** is an extension of the signature of **Lingua** (Sec.5.7) by new constructors. The carriers change due to new SQL-values and SQL-types.

## 6.8   Denotations and their constructors

The subalgebra of expression denotations of all types is analogous as in **Lingua**.

At the level of state-to-state functions we have a new domain of transactions. *Transactions*, similarly to instructions, are state transformations but contrary to the former they are total functions since they do not contain loops and procedure calls. Moreover, they do not create new tables but only modify the existing ones. Their domain is, therefore, the following:

trd : TrnDen = State $\mapsto$ State

Transactions are regarded as a separate carrier of our the algebra of denotations to avoid the use of arbitrary table instructions in the contexts of transactions.

The largest group of transactions are *table modifications* which in a traditional syntax could have the form:

```
ide := table-expression(ide)
```

where on both sides we have the same table named `ide`. Transactions include the mechanisms of creating and recovering security copies of databases.

The carrier of instruction denotations is enriched with new constructors of specific SQL instructions of three categories;

1. row assignments,
2. table assignments,
3. database instructions.

All constructors of **Lingua** are still available and apply to the extended carrier of instruction denotations. This rule concerns, in particular, the constructor of transfer replacement and the constructors of structural instruction, i.e., sequential composition, branching and loop. The constructors of procedure declaration and procedure call remain unchanged as well, although now they are defined on extended domains.

A particular role in SQL plays a large group of table assignments where we distinguish two categories:

1. *table-modification instruction* where on both sides of the assignment we have the name of the same table; this group of instructions comprise the mechanisms known as CASCADE and RESTRICT,
2. *table-creation instruction* where on the left-hand side of the instruction we may have a different table name (of the table that is being created) than on the right-hand side.

From a mathematical perspective the first category may be regarded as a particular case of the second, but denotationally they correspond to two different constructors of the algebra of denotations hence also to different constructors of the algebra of syntax.

Independently of the described categorisation, table assignments are split into two further categories according to two ways of using subordination constraints both described in Sec. 11.5 of [16][16]):

1. *conformist instructions* where an execution terminates with an error message whenever it would lead to a violation of subordination constraints; this category corresponds to the option RESTRICT,

2. *correcting instructions* which in the described situation introduce such changes into a database that guarantee the protection of subordination constraints; this category corresponds to the option CASCADE.

*Queries* are similar to simple instructions with the difference that they always create a new table assigned to the system-identifier monitor. Consequently, we apply simplified assignments as-sign-mo that never violates any constraints since the transfer of the new value is TT.

*Cursors* are mechanisms used to get row-by-row from tables. In our model that can be easily defined, e.g. by adding a column to a table that enumerates its rows.

*Views* are essentially procedures that call table instructions. They may be introduced to our model either as predefined instructions or by providing programming mechanisms of procedures that operate on tables.

Regarding data base instructions I assume that in **Lingua-SQL** an initial valuation of program execution may carry some variables assigned to database values. I assume additionally that in every initial state of program execution, the system identifiers are bound to the following default values:

tye.sb-graph = Ø

vat.copies = Ø,

vat.monitor = Ω  (interpreted as no data to be displayed)

vat.check = 'yes'

With these assumptions each database program in **Lingua-SQL** that operates on tables either has to create its own tables — and a database thereof — or to activate an already existing database. In **Lingua-SQL** we have therefore only two database instructions that operate on tables — activate and archive — and two that operate on subordination graphs, which add or remove an edge of a graph.

## 6.9  An example of a colloquial syntax

The colloquial syntax of **Lingua-SQL** should be as close as possible to SQL standard. Below just one example of restoring a standard table-variable declaration — which in **Lingua-**SQL belongs to colloquial syntax — into its corresponding concrete-syntax form.

```
create table Employees with
  Name            Varchar(20)   NOT NULL,
  Position        Varchar(9),
  Salary          Number(5)     DEFAULT 0,
  Bonus           Number(4)     DEFAULT 0,
  Department_Id   Number(3)     REFERENCES Departments,
  CHECK (Bonus < Salary)
ed
```

The restoring transformation would change this declaration into a sequence of a table-variable declaration followed by a database instruction of retting a subordination dependency between tables:

   **create table** `Employees` **as** typ_exp **ed** ;

   **set reference of** `Employees` **et** `Department_Id` **to** `Departments` **ei**

where typ_exp is a metavariable that represents a type expression:

   **table-type** dat_exp **with** tra_exp **ee**

In this scheme the data expression dat_exp defines data that stand in the row of default data which in fact means that it generates this row. In turn the transfer expression tra_exp describes the properties of columns and rows. The table-variable declaration has then the form:

   **create table** `Employees` **as**

     **table-type** dat_exp **with** tra_exp **ee**

   **ed**

Unfolding the data expression by means of row-creation and row-expansion constructors and unfolding the transfer expression with transfer-expression constructors we get the following concrete version of our colloquial declaration:

   **create table** `Employees` **as**                    (the beginning of the declaration)

     **table-type**                                   (the beginning of type expression)

       **expand-row**                                (the beginning of data expression)

         **expand-row**

           **expand-row**

             **expand-row**

               **row** `Name` **val** **empty-word ee**

            **by** `Position` **val empty-word ee**

         **by** `Salary` **val** 0 **ee**

       **by** `Bonus` **val** 0 **ee**

     **by** `Department_Id` **by empty-number ee**     (the end of data expression)

     **with**                     (the beginning of transfer expression yoke expression)

       **all**

         `varchar(20)(`**row**`.Name)`         **and**

         `not-null(`**row**`.Name)`            **and**

         `varchar(9)(`**row**`.Position)`       **and**

         `number(5)(`**row**`.Salary)`         **and**

         `number(4)(`**row**`.Bonus)`          **and**

         `number(3)(`**row**`.Department_Id)`   **and**

```
        row.Bonus < row.Salary
    ee                                  (the end of transfer expression (yoke expression)
  ee                                                    (the end of type expression)
 ed ;                                                        (the end of declaration)
set reference of Employees et Department_Id to Departments ei
```

Of course `varchar(20), varchar(9),`… are the names of appropriate predicates. Notice that in this example one "syntax unite" from the colloquial level is transformed into a sequential composition of a declaration with an instruction.

## 6.10   Remarks about a possible implementation of Lingua-SQL

Typical *Application Programming Interfaces* (API) for SQL have been created for programming languages such as e.g. C, PHP, Perl, and Phyton. Each of these programming environments constitutes a programming language equipped with the mechanisms that allow to run procedures of a certain existing database-engine. In the case of **Lingua-SQL,** such a situation would not be acceptable. Our language must be based on a dedicated SQL-engine with a denotational model, and in the future, maybe, with a dedicated implementation. Such an approach is necessary, if we want to provide sound program-construction rules for **Lingua-SQL**.

# 7   What remains to be done

Even though [16] is already of a considerable size, the majority of subjects has been only sketched. Below a preliminary list of subjects which could be developed further. This list is certainly not complete.

## 7.1   The development of Lingua

1. **An extension of Lingua** to some "practical" language, say **Lingua-α**, where preliminary programming experiments could be performed. Such a language should cover in particular:

    1.1. The mechanisms of object programming which in [16] have been only sketched in Sec.9.

    1.2. Some more specific data types, e.g. trees that in the Polish version of [16] have been sketched in Annex 1.

    1.3. The enrichment of SQL mechanisms.

    1.4. The elaboration of HTML scripts.

2. **The development of tools for correct-programs' development in Lingua-α:**

    2.1. The extension of the languages of conditions and thesis sketched in Sec.8 of [16].

    2.2. Sound program-construction rules for the extended language.

3. **A user manual for Lingua-α.** This task could also contribute to a methodology of writing programmer's manuals for languages with denotational semantics[29].

---

[29] Denotational models should provide an opportunity for the revision of current practices seen in the manuals of programming languages. New practices should on one hand base on denotational models

4.  **A programmer's environment for Lingua-α:**

    4.1. An interpreter or a compiler. To make this interpreter/compiler maximally independent of possible errors in the language used to build it, some basic core could be coded in such a language (e.g. in Python), and the remaining part may be written using this basic core. This could also be the first experiment in using our language.

    4.2. An editor of programs supporting the construction of correct programs with the use of earlier developed program construction rules (see 2.2)

    4.3. An adaptation of an existing theorem prover for proving metaconditions (the properties of conditions) described in Sec. 8.4.2 of [16] which is necessary for the use of program-construction rules.

5.  **Preliminary experiments with programming in Lingua-α**:

    5.1. Microprograms due to their relatively small volume and a very critical correctness issue.

    5.2. Simple SQL applications due to the availability of tools.

This is, of course, only a preliminary sketch of a project which — in the case of realizations — would probably be modified and further developed.

## 7.2 The development of a software environment for language designers

Such an environment should consist of:

1. An editor of the definitions of denotations' constructors.
2. A generator of the grammar of abstract syntax from such definitions.
3. An editor supporting language designers in developing concrete-syntax grammar from abstract syntax grammars.
4. An editor/generator of a transformation restoring colloquial syntax to abstract syntax.
5. A generator of a parser from colloquial syntax to abstract syntax.
6. A generator of an interpreter of the language.

If such an environment is created before **Lingua-α**, it could be used in the creation of that language.

## 7.3 Two basic research problems

Independently of the tasks mentioned above, two important research problems are worthy of consideration.

---

but on the other — do not assume that todays' readers are acquainted with it. A manual should provide some basic knowledge and notation needed to understand the definition of a programming language written in a new style. At the same time — I strongly believe on that — it should be written for professional programmers rather than for amateurs. The role of a manual is not to teach the skills of programming. Such textbooks are, of course, necessary, but they should tell the readers what the programming is about rather than the technicalities of a concrete language. An experiment in writing a user manual of **Lingua** is described in [15].

The first concerns the extension of our model by the mechanisms of concurrency. Fully denotational models of concurrence are not known today, although there are some attempts to "semi-denotational" models of these mechanisms, as, e.g. in [2].

The second problem has not been probably tackled at all and concerns the construction of semi-formal languages for the description of user-oriented specifications of programs. So far all approaches to program correctness — including mine — concentrate on the compatibility of program code with its formal specification. It does not exhaust the reliability problem in the IT industry, because many problems are due to poor communication between a designer of a system and its user. Most probably many area-oriented languages of specifications would be needed.

# 9 Index